\documentclass[preprint,twocolumn]{aastex6}
\usepackage{natbib,amsmath,multirow}

\newcommand{\bjdtdb}{\ensuremath{\rm {BJD_{TDB}}}}
\newcommand{\feh}{\ensuremath{\left[{\rm Fe}/{\rm H}\right]}}

\newcommand{\msun}{\ensuremath{\,M_\Sun}}
\newcommand{\rsun}{\ensuremath{\,R_\Sun}}
\newcommand{\lsun}{\ensuremath{\,L_\Sun}}
\newcommand{\mj}{\ensuremath{\,M_{\rm J}}}
\newcommand{\rj}{\ensuremath{\,R_{\rm J}}}

\newcommand{\fave}{\langle F \rangle}
\newcommand{\fluxcgs}{10$^9$ erg s$^{-1}$ cm$^{-2}$}

\newcommand{\teff}[1]{\ensuremath{T_{\rm eff,#1}}}
\newcommand{\fbol}{\ensuremath{F_{\rm bol}}}
\newcommand{\sig}[1]{\sigma_{#1}/#1}
\newcommand{\sigsb}{\ensuremath{\sigma_{SB}}}

\newcommand{\Sig}[1]{\left(\frac{\sigma_{#1}}{#1}\right)}

\newcommand{\del}{\partial}
\newcommand{\logg}{\ensuremath{\log(g_1)}}

\begin{document}
\title{Measuring Model-independent Masses and Radii of Single-Lined Eclipsing Binaries: Analytic Precision Estimates}
\author{Daniel J. Stevens\altaffilmark{1}, B. Scott Gaudi\altaffilmark{1}, and Keivan G. Stassun\altaffilmark{2,3}}
\altaffiltext{1}{Department of Astronomy, The Ohio State University, 140 W. 18th
Avenue, Columbus, OH, USA 43210}
\altaffiltext{2}{Vanderbilt University, Department of Physics \& Astronomy, 6301 Stevenson Center Ln., Nashville, TN 37235, USA}
\altaffiltext{3}{Fisk University, Department of Physics, 1000 17th Ave.\ N., Nashville, TN 37208, USA}

\begin{abstract}
We derive analytic estimates for the ability with which one can obtain precise, empirical stellar masses and radii via single-lined eclipsing binaries (EBs) in the era of {\it Gaia\/} and {\it TESS}.  Including stars that host transiting substellar companions, such single-lined EBs already number in the hundreds from ground-based transit surveys and will comprise a major component of the science yield from the upcoming {\it TESS\/} mission.
We explore the requirements for obtaining a given fractional precision on the masses and radii of single-lined EBs using primarily empirical means: radial velocity and eclipse measurements along with either: estimates of the primary's (1) surface gravity from high-resolution spectroscopy; (2) radius inferred from parallax, effective temperature, and bolometric flux; or (3) surface gravity and density from asteroseismology. We then compare these requirements to the precision obtained from invoking stellar models or empirical relations. We show that, for a fiducial transiting hot Jupiter system, precise, accurate, and essentially model-independent mass and radius measurements for such single-lined EBs will be possible in the era of \emph{Gaia}.  These will be comparable in precision to those obtained with double-lined EBs. Moreover, the systems for which these methods can be applied will vastly outnumber double-lined EBs, thereby possessing the potential to sample a more complete range of stellar types (such as M dwarfs); these systems will also, in many cases, be more amenable to precision metallicity and abundance determinations than are double-lined EBs. 
\end{abstract}

\section{Introduction}\label{sec:intro}

Measurements of stellar and planetary masses and radii are paramount to our understanding of stellar and planetary evolution. Different physical prescriptions in stellar evolution models for various effects (e.g., mass loss, atmospheric boundary conditions, convective overshoot and mixing length, rotation, and mass loss due to winds, just to name a few) predict different masses and radii for stars of the same mass, age, and metallicity.  Similarly, stars with different elemental abundance ratios will have significantly different evolutionary paths in the luminosity-effective temperature plane, even if they have the same mass and overall metal abundance.   Thus, placing precise constraints on these parameters is our only method of empirically constraining the wide range of plausible stellar evolution models. The properties of exoplanets discovered via the radial velocity (RV) and transit methods are only known relative to the properties of the host star, so it is also imperative to characterize stars precisely and accurately to calibrate stellar models and to study planetary composition, formation, and evolution.

 Measured K- and M-dwarf radii have been shown to exceed model-predicted radii at fixed mass by 5-10\%, and their effective temperatures are suppressed by similar amounts, or $\sim 150$ K. Analyses of high-precision data on the low-mass eclipsing binaries (EBs) YY Geminorum (\citealt{Torres2002}), CU Cancri (\citealt{Ribas2003}), GU Bootis \citep{Lopez-Morales2005}, and CM Draconis \citep{Morales2009} were among the first to notice these discrepancies, while subsequent analyses have found these discrepancies to be endemic to low-mass stars  (cf. \citealt{Kraus2011,Birkby2012}). This radius discrepancy is often referred to in the literature as the ``radius inflation'' problem, although the observed radii are ostensibly the true values, and the models are underestimating them. The K- and M-dwarf radius discrepancy has yet to be fully captured in stellar models \citep[see, e.g.,][and references therein]{Stassun:2012,Somers:2016}.
 
 In the case of binaries, there is evidence that increased stellar activity due to tidal interactions \citep{Lopez-Morales2007} may be responsible through increased star-spot coverage and/or increased magnetic field strength and a correspondingly stronger inhibition of convective heat transport \citep{Feiden2013}. Magnetic stellar evolution models can produce surface field strengths that agree with those inferred from observations of, e.g. X-ray flux \citep{MacDonald2014}, but it is unclear whether the consequently large interior magnetic field strengths are physical (e.g. \citealt{Feiden2014}). A potential missing opacity source in the stellar models has been discussed as a cause as well (\citealt{Berger2006,Lopez-Morales2007}). Conversely, \citet{Mann2015} found no association between the radius discrepancy and either metallicity or standard tracers of magnetic activity.
 
 If tidal interactions were responsible for the observed radius and effective temperature discrepancies, however, one would expect the discrepancy to disappear for M dwarfs in longer-period binaries, where the tidal forces are weaker; the 41-day period M-M binary LSPM J1112+7626 \citep{Irwin2011} also shows these discrepancies. As this is only one long-period low-mass EB, there is significant motivation to characterize others precisely and accurately in order to determine if the radius and effective temperature anomalies persist out to large separations at which the individual binary components might be expected to behave like isolated stars.
 
 Resolving these tensions is critical not only for our understanding of low-mass stellar evolution but also for our understanding of the planets that orbit such stars. The M dwarfs are prime targets for exoplanet searches, given that the "habitable zone" lies closer to an M dwarf than it does for hotter stars, and the stars themselves are less massive and smaller; thus, it is easier to detect terrestrial planets in the habitable zones of M dwarfs via the transit and RV methods.

For these reasons, low-mass stars make excellent targets for transit surveys for habitable planets \citep{Gould2003b}, such as MEarth \citep{Nutzman2008} and TRAPPIST \citep{Jehin2011}.  Because the depths of the transits are relatively large, the ratio of the radius of the planet to the star can be measured to relatively high precision -- e.g. to $<5\%$ for GJ1132b \citep{BertaThompson2015} and to $0.5-5\%$ for the TRAPPIST-1 planets \citep{Gillon2017} -- meaning that a $5-10\%$ inaccuracy in the M-dwarf radius will contribute significantly to the precision of the inferred planet's radius. Additionally, M dwarfs are prone to violent events such as flares that affect an orbiting planet's habitability, and these events depend on the poorly understood magnetic activity of the star. We therefore need to characterize these low-mass host stars accurately and precisely both to determine the physical properties of the planet to the necessary precision and to investigate their habitability.

For higher-mass host stars, there exists the "retired A stars" controversy: \citet{Johnson2007} targeted subgiants and giants with masses $\sim 1.5\msun$ (i.e., "retired A stars") to search for planets, but \citet{Lloyd2011} suggested that the majority of these stars were not, in fact, retired A stars, arguing instead that the masses lie between $1-1.2\msun$. This has led to a robust debate in the literature, as efforts have been made to determine the true masses of these stars with simulations (\citealt{Johnson2013,Lloyd2013}), spectroscopy and asteroseismology (\citealt{Johnson2014,Ghezzi2015,Campante2017,North2017,Stello2017}), and a combination of parallaxes, spectral energy distributions (SEDs), RVs, and transits \citep{StassunGaiaPlanets:2016}.

Up to this point, precise ($\lesssim 3\%$ fractional uncertainty) empirical measurements of both stellar masses and radii have been largely restricted to double-lined EBs and the $\alpha$ Centauri system. As a result of Malmquist bias and photon noise, the stars in these binaries tend to be bright and, as a result, fairly high mass. \cite{Torres2010} used 95 such systems to derive mass-radius relations, but the low-mass end of the sample is, consequently, sparsely populated: only nine of the 190 stars are M dwarfs, and only two of those have masses below the fully convective boundary at $M \lesssim 0.3 M_{\sun}$. A couple dozen M-M binaries in the literature have parameters measured to $<5\%$ precision (see \citealt{Lubin2017}), though getting accurate parameters is a challenging task (e.g. \citealt{Irwin2011}).

Single-lined EBs can also provide masses and radii, but without the RV orbit of the companion, one can only measure combinations of the component masses and radii. These systems typically require the use of stellar evolution models or empirical relations to break these degeneracies. The use of stellar models or empirical relations implicitly assumes that the stars being analyzed are typical of the stars used to calibrate the models and relations. As such, the parameters inferred for such stars tell us nothing new about how the fundamental physical stellar parameters relate to each other, and they may be inaccurate if the analyzed stars are atypical. To validate existing models or to learn more about the relations between stellar parameters, we need to determine these quantities in a model-independent way; here, ``model-independent'' means including neither stellar evolution models nor empirically calibrated relations in the analysis, as these are precisely the models and relations we need to refine.  Alternatively, tidal synchronization has been invoked for sufficiently short orbital periods -- e.g. \citealt{Fernandez2009} -- to obtain the primary stellar radius independently. However, \citet{DittmanEB2017} discovered a low-mass EB on a $\sim4.7$ day orbit that is not tidally synchronized, so this assumption is clearly not always valid at low periods. Additionally, one cannot determine {\it a priori} for which systems it is valid.

Exoplanet transit surveys have already identified a wealth of single-lined EBs -- both false positives (e.g. \citealt{Collins2018} and the exoplanetary systems themselves. However, it is impossible to measure the masses and radii of each binary component independently from only the photometric light curve and RVs; instead, one measures only the primary density (if one has a constraint on the eccentricity of the system or if one can safely assume that the orbit is nearly circular) and the secondary's surface gravity \citep{Seager2003}. These quantities are insufficient to break the degeneracy between the radius and mass of the primary, and thus determine these properties for the secondary. High-resolution spectroscopy of the primary star can provide the primary's surface gravity, thus in principle allowing one to break this degeneracy, but the uncertainties are often of order tenths of a dex -- and often worse for hotter, rapidly rotating stars with few, broad spectral lines.

Parallax measurements can provide additional constraints on the primary star's parameters. With a distance determined from parallax, along with an effective temperature and a bolometric flux from an SED, one can calculate the stellar radius, thus breaking the aforementioned degeneracies. This requires precise and accurate parallaxes, which {\it Gaia} \citep{Gaia_main} will provide at the end of its mission: parallaxes with uncertainties as small as $5-10\mu$as are expected for the brightest, most nearby stars. These uncertainties translate to distances with sub-percent precision. As an example, \citet{Stevens2017} used parameterized SEDs and {\it Gaia} Data Release 1 (DR1) parallaxes \citep{GaiaDR1} to calculate radii for over 350,000 stars to a median precision of a few percent.

\citet{StassunGaiaPlanets:2016} applied this method to $\sim 500$ exoplanet host stars, using SED-derived angular diameters and {\it Gaia\/} DR1 parallaxes to measure accurate and empirical stellar and planetary parameters. With the retired A star hot Saturn system KELT-11, \citet{Beatty2017} employed this method to show how the exquisite parallaxes expected from {\it Gaia} will allow for precise, model-independent mass and radius measurements that can be used to test empirical relations -- and, in this case, to determine whether or not KELT-11 is indeed a retired A star.

Furthermore, precise time-series photometry from surveys such as {\it Kepler} and the {\it Transiting\ Exoplanets\ Survey\ Satellite}\ ({\it TESS}; \citealt{Ricker:2015}) can provide both the primary mass and radius via asteroseismology. {\it TESS} will deliver short-cadence light curves over a continuous span of $\sim 27-300$ days for hundreds of thousands of stars and single-lined EBs  \citep{Sullivan2015,Sullivan2017}. For some of these systems -- particularly the stars closer to the ecliptic poles, which will be observed for the longest amount of time \citep{Creevey2017} -- it should be possible to measure the global oscillation properties: the large frequency spacing $\delta\nu$ which is directly related to the bulk density of the star, and the frequency of maximum power $\nu_{max}$, which provides the stellar surface density via an empirically calibrated scaling relation \citep{Kjeldsen1995}.  These can then be combined to estimate the mass and radius of the star. In the case of the stars with the best photometric data, it may also be possible to improve on these estimates by "peak-bagging" and modeling individual frequency peaks (e.g. \citealt{Deheuvels2012,Huber2013,Huber2013Sci, Metcalfe2014,Lund2017,Silva2017}) or the ratios of frequencies \citep{Creevey2017}, thereby improving the uncertainty on these parameters. This analysis should be applicable to around 200 planet hosts either recovered or discovered by {\it TESS} \citep{Campante:2016}. For the stars closer to the ecliptic plane that will be observed for shorter amounts of time, it may be possible to apply empirical corrections (\citealt{Kjeldsen2008,Ball2014,Ball2014err}) to improve the accuracy of the modeling \citep{Schmitt2015}.

Finally, high-precision, high-cadence time-series photometry such as what {\it TESS} will provide also allows one to measure granulation-induced brightness modulations such as the light-curve ``flicker,'' which is due to stellar granulation and has been shown to correlate well with the stellar surface gravity (\citealt{Bastien2013, Bastien2016}). Measuring the timescale of the granulation-driven modulations and the $p-$mode oscillations can also provide a precise constraint on the surface gravity (\citealt{Kallinger2014,Kallinger2016}). These techniques thus present another means by which to break the mass/radius degeneracy \citep[see][for a detailed discussion of this approach in the context of {\it Gaia\/} and {\it TESS\/}]{StassunGrav:2017}.

All together, these techniques and upcoming data sets promise to elevate single-lined EBs as true stellar benchmarks that are superior in some respects to double-lined EBs. The present work lays out the methodology to fully harness this potential. In Section \ref{sec:setup}, we define the physical parameters of interest for a single-lined eclipsing system. In Section \ref{sec:obs}, we explore the corresponding observable quantities for transits/eclipses and RVs. We derive expressions for the component masses and radii in terms of these observables and external constraints from parallax, asteroseismology, spectroscopy, empirical relations, and isochrones in Section \ref{sec:scale}, and we derive analytic estimates of the precision needed to attain $3\%$ masses and radii in Section \ref{sec:precision}. We perform an analysis on synthetic data of a model hot Jupiter system in Section \ref{sec:HJ} and discuss the effects of eccentricity and limb darkening to the error budgets in Section \ref{sec:caveats} . We conclude by discussing the anticipated number of systems for which precision parameters will be obtainable in Section \ref{sec:yield}.

\section{Methods}\label{sec:methods}

\subsection{Problem Set-Up}\label{sec:setup}

Consider a binary system with masses $M_1$ and $M_2$, radii $R_1$ and $R_2$, semimajor axis $a$, period $P$, and inclination $i$. Let $\rho_1$ ($\rho_2$), $g_1$ ($g_2$), and $\teff{1}$ ($\teff{2}$) refer to the density, surface gravity, and effective temperature of the primary (secondary). Let $q \equiv M_2/M_1$ and $k \equiv R_2/R_1$ be the mass and radius ratios of the system. We will consider the case in which $q \ll 1$ and $k \ll 1$, i.e. systems consisting of a stellar primary that is of solar type or earlier that is orbited by a planetary or low-mass stellar companion.

We also assume that the companion is relatively ``dark,'' i.e., that it contributes negligibly to the total flux of the system. Given the goal of achieving percent-level accuracy and precision on the masses and radii, this requires the companion to contribute less than 1\% of the total flux: $F_2/F_1 \lesssim 0.01$, where $F_1$ and $F_2$ are the fluxes of the primary and secondary binary components, respectively. Adopting the mass-bolometric luminosity relation of the stars with $0.4 \leq M/M_{\sun} \leq 1.5$ in the \citet{Torres2010} sample of well-vetted binaries, $M/M_{\odot} \simeq (L/L_{\odot})^{0.2}$. Thus, $q \equiv M_2/M_1 \sim (F_2/F_1)^{0.2}$, since both binary components are effectively at the same distance from the observer. Our $<1\%$ flux ratio requirement necessitates $q \lesssim 0.4$, or $M_2 \lesssim 0.4M_1$.

This relationship holds for most M dwarfs orbiting A-, F-, or G-dwarf primaries, and as such, our assumption of a dark companion is then reasonable for a broad range of EBs ($\sim 35\%$ of binaries, according to the \citealt{Raghavan2010} mass ratio distribution). We note that the exact flux ratio depends on the wavelength range under consideration: the companion will contribute a larger fraction of the total flux at wavelengths where both stars are in the Rayleigh-Jeans tail of the SED, since, in this regime, the flux ratio is proportional to the effective temperature ratio. This means that the flux ratio will be higher in, e.g., the near-infrared $J,\ H,\ {\rm and}\ K$ filters than in the optical filters. As such, for a robust analysis of an EB, the dark-companion assumption needs to be validated on a filter-by-filter basis for both the transit/eclipse photometry and the broadband flux measurements used in the SED fits. 

Following Equation 21 of \citet{Winn2010}, the observed flux at a given time during the transit (total flux of both bodies minus the flux obscured by the transit) depends only linearly on the companion-to-primary flux ratio; the observed flux has an additional dependence on the product of the square of the radius ratio and a dimensionless constant of order unity that describes the geometric overlap of the binary disks during transit.

We further restrict our binary to a circular orbit (such that the eccentricity $e=0$) to make the analysis more analytically tractable. We consider the effect of eccentricity on the error budget in Section \ref{sec:caveats}. Additionally, we require that the impact parameter $b = (a/R_1)\cos i < 1-k$: the transit observables defined below are appropriate for a light curve that is piecewise linear in time, which is increasingly inaccurate when describing near- and fully grazing eclipses \citep{Carter2008}. Given the small values of $q$, $k$, and $F_2/F_1$, the system under consideration will generally be a single-lined eclipsing binary for which typically only the primary eclipse will be easily visible.

\subsection{Observable Quantities}\label{sec:obs}
Since the primary eclipse light curve only gives $\rho_1$ (for $k \ll 1$ and a circular orbit) and RV only gives $M_2/M_1^{2/3}$ (for $M_2 \ll M_1$), we need additional information to back out the mass and radius for each component. We show these results below.

\subsubsection{\label{sec:LCO}Light-curve Observables}
The light curve of a single-lined EB can be described by the total flux of the system, the eclipse duration, the ingress/egress duration, the period, the transit depth, and the limb darkening. To make the problem analytically tractable, we will first assume no limb darkening and approximate the transit as a trapezoidal shape; we explore limb-darkening effects in Section \ref{sec:caveats}. We also assume that no transmission of the primary's light through the companion's atmosphere occurs when the companion passes in front of the primary. Finally, we assume that each flux measurement is normalized to the median out-of-eclipse value, i.e. that $f_0 =1$. In this case, the observables are the period $P$, the FWHM $T$ of the transit, the ingress/egress duration $\tau$, the transit depth $\delta$, and the time of transit center $T_c$ \citep{Carter2008}.

If we define the diameter crossing time $\tau_0 \equiv R_1P/(2\pi a)$, then the aforementioned observables are related to the physical parameters $\tau_0$, $b$, and $k$ by

\begin{equation}
T \equiv 2\tau_0 \sqrt{1-b^2}
\end{equation}
\begin{equation}
\tau \equiv 2\tau_0 \frac{k}{\sqrt{1-b^2}}
\end{equation}
and
\begin{equation}
\delta \equiv k^2,
\end{equation}

and the inverse mappings are given by
\begin{equation}
b = \sqrt{1 - k\frac{T}{\tau}},
\end{equation}
\begin{equation}
\tau_0 = \sqrt{\frac{T\tau}{4k}},
\end{equation}
and
\begin{equation}
k = \sqrt{\delta}
\end{equation}.

The diameter crossing time as defined above assumes that, during transit, the transiting body's orbital velocity vector lies entirely within the plane of the sky -- i.e. that it has no radial component. In reality, as the companion moves across the face of the primary star, it travels in an arc, such that the sky-projected velocity is less than the orbital velocity. For simplicity, we approximate that arc as a chord across the primary stellar disk such that the orbital and sky-projected velocities are equal.

From this light-curve model, it is impossible to determine the absolute scale (e.g. semimajor axis) of the system; rather, only dimensionless quantities, such as $\delta$, $\tau/P$, $T/P$, and $T_c/P$ are measurable. However, one can infer
\begin{equation}
\label{eq:a/R*} a/R_1 = P/(2\pi\tau_0)
\end{equation}
for a circular orbit \citep{Seager2003}. With this scaling, one can match the observables with eclipsing systems of arbitrary size, whether or not the inferred parameters are physically plausible. This scaling, combined with Kepler's third law of planetary motion, enables one to measure a combination of the densities \citep{Seager2003}:
\begin{equation}
\label{eq:densityeq} \rho_1 + k^3\rho_2 = \frac{3\pi}{GP^2}\left(\frac{a}{R_1}\right)^3,
\end{equation}
where $G$ is the gravitational constant. If $k \ll 1$, then one can infer the density of the primary from the light curve. Hence, if one can measure a different combination of the primary's mass and radius, then it is possible to break these degeneracies and determine the absolute scale of the system. We discuss various ways of doing this below.

\subsubsection{\label{sec:RVO}RV Observables}
From a single-lined spectroscopic binary (SB1), the RV observables are the period $P$, the primary star's RV semi-amplitude $K_1$, eccentricity $e$, argument of periastron $\omega$, and time of periastron (or some other reference time in the orbit, such as the time of primary eclipse). As is well known, these observables do not allow for the unique determination of individual masses; rather, they enable one to infer the mass function of the system,
\begin{equation}
\label{eq:massfunc} {\cal M} = \frac{M_2 \sin i}{(M_1 + M_2)^{2/3}},
\end{equation}
since
\begin{equation}
\label{eq:RV} K_1 = (2\pi G)^{1/3}P^{-1/3}{\cal M}.
\end{equation}
If $q \ll 1$ and one assumes a value for $M_1$, then one can deduce $M_2 \sin i$ -- the minimum mass of the secondary. The companion mass is then given by
\begin{equation}
\label{eq:m2rv} M_2 = (2\pi G)^{-1/3} \frac{K_1P^{1/3}}{\sin i}M_1^{2/3}.
\end{equation}
If the system is also eclipsing, then $\sin i \approx 1$ and one can infer the true mass of the secondary, given the assumption about $M_1$. Hence, even for an eclipsing SB1, one cannot determine the absolute scale of the system, and combinations of the primary mass and radius from other measurements are necessary.

It is worth noting that, with only eclipse and RV measurements, one can determine the surface gravity of the secondary using only directly observable quantities:
\begin{equation}
g_2 = \frac{2\pi}{P}\frac{K_1}{(R_2/a)^2\sin i}.
\end{equation}

\subsection{Mapping Observables to Masses and Radii}\label{sec:scale}
\subsubsection{\label{subsec:specscale}Photometric $\rho_1$ and Spectroscopic $\log(g_1)$}

Since $10^{\log(g_1)} = GM_1/R_1^2$ and $\rho_1 = 3M_1/(4\pi R_1)^3$, if we can estimate \logg\ (via, e.g., high-resolution spectra or flicker), we immediately recover the primary's mass and radius:
\begin{equation}
\label{eq:m1gdens}
M_1 = \frac{9}{16 \pi^2 G^3}\frac{g_1^3}{\rho_1^2}
\end{equation}
and
\begin{equation}
\label{eq:r1gdens}
R_1 = \frac{3}{4\pi G} \frac{g_1}{\rho_1}.
\end{equation}
The companion's radius also follows:
\begin{equation}
\label{eq:r2gdens}
R_2 = \frac{3}{4\pi G} \frac{g_1}{\rho_1} \delta^{1/2}.
\end{equation}
The RV measurements give us, via Equation \ref{eq:RV},
\begin{equation}
\label{eq:m2gdens}
    M_2 = \left(\frac{81}{512}\right)^{1/3}\pi^{-5/3}G^{-7/3} \frac{K_1P^{1/3}}{\sin i}\frac{g_1^2}{\rho_1^{4/3}}.
\end{equation}
From the above, we see that the inclination only explicitly affects the secondary's mass.

The primary density $\rho_1$ can be written in terms of light-curve observables by combining Equations \eqref{eq:a/R*} and \eqref{eq:densityeq}:
\begin{equation}
\label{eq:density} \rho_1 \approx \frac{3 P}{8\pi^2G}\tau_0^{-3} = \frac{3 P}{\pi^2G}(T\tau)^{-3/2}\delta^{3/4}.
\end{equation}
We can then express the masses and radii in terms of observables:
\begin{equation}
\label{eq:m1g}
M_1 = \frac{\pi^2}{16G}\frac{(g_1T\tau)^3}{P^2\delta^{3/2}}
\end{equation}
\begin{equation}
\label{eq:r1g}
R_1 = \frac{\pi}{4}\frac{g_1(T\tau)^{3/2}}{P\delta^{3/4}}
\end{equation}
\begin{equation}
\label{eq:m2g}
M_2 = \frac{\pi}{512^{1/3}G}\frac{K_1g_1^2}{P\delta}(T\tau)^2
\end{equation}
\begin{equation}
\label{eq:r2g}
R_2 = \frac{\pi}{4}\frac{g_1 (T\tau)^{3/2}}{P\delta^{1/4}}.
\end{equation}
From these relations, we see that the error on these four parameters will depend strongly on the primary's surface gravity, the orbital period, the FWHM eclipse duration, and the ingress/egress duration. Section \ref{sec:specsig} details our analytic error estimates.

\subsubsection{\label{subsec:parallaxscale}Photometric Density and Parallax}
A measurement of the star's parallax angle in radians, $\pi_p$, gives us the distance to the star in AU, $d_p \approx {\rm 1\ AU}/\pi_p$. If the primary star's SED is sufficiently sampled to estimate its bolometric flux $\fbol$, and if we have an estimate of the primary's effective temperature $\teff{1}$ (from, for example, fitting a template spectrum to high-resolution spectrum or from the SED itself), we can calculate $R_1$:

\begin{equation}
\label{eq:r1pphys} R_1 = \frac{1\ \rm AU}{\pi_p}\frac{\fbol^{1/2}}{\sigsb^{1/2}\teff{1}^2},
\end{equation}
where \sigsb\ is the Stefan-Boltzmann constant. It then follows from the eclipse photometry that
\begin{equation}
\label{eq:m1pphys} M_1 = \frac{4\pi}{3}\rho_1R_1^3 = \frac{4\pi}{3}\rho_1 \left(\frac{1\ \rm AU}{\pi_p}\right)^{3}\frac{\fbol^{3/2}}{\sigsb^{3/2}\teff{1}^{6}}
\end{equation}
and
\begin{equation}
\label{eq:r2pphys} R_2 = kR_1 = \frac{1\ \rm AU}{\pi_p}\delta^{1/2}\frac{\fbol^{1/2}}{\sigsb^{3/2}\teff{1}^{2}}.
\end{equation}
Finally, via RV observations and Equation \eqref{eq:m2rv},
\begin{equation}
\label{eq:m2pphys} M_2 = \left(\frac{8\pi}{9G}\right)^{1/3} \frac{K_1}{\sin i}P^{1/3}\rho_1^{2/3}\left(\frac{1\ \rm AU}{\pi_p}\right)^{2}\frac{\fbol}{\sigsb\teff{1}^{4}}.
\end{equation}
In terms of eclipse observables, Equations \ref{eq:m1pphys} and \ref{eq:m2pphys} can be expressed as
\begin{equation}
\label{eq:m1p} M_1 = \frac{4}{\pi G}P \left(\frac{\fbol}{\sigsb T\tau}\right)^{3/2}\delta^{3/4}\left(\frac{\rm 1\ AU}{\pi_p}\right)^{3}\teff{1}^{-6}
\end{equation}
and
\begin{equation}
\label{eq:m2p} M_2 = \frac{2}{\pi G}\frac{K_1P}{\sin i}\frac{\fbol}{\sigsb\teff{1}^{4}}\left(\frac{\rm 1\ AU}{\pi_p}\right)^2 \frac{\delta^{1/2}}{T\tau}.
\end{equation}

This analysis assumes that the measured parallax is the true parallax, but this is not necessarily true: for example, Lutz-Kelker bias \citep{Lutz1973} would make one more likely to underestimate the distance to a particular star. This would result in an underestimate of the stellar luminosity, as well as of the stellar radius for fixed \teff.

For some stars, it is possible to measure their angular diameters $\Theta_D \equiv 2R_1/d_p$ interferometrically. For those interferometric stars that have trigonometric parallaxes (and hence distances), these two measurements give the primary radius with no need for SED fitting. In this case, linear error propagation, ignoring covariances, gives $(\sig{R_1})^2 \approx (\sig{\Theta_D})^2 + (\sig{\pi_p})^2$, since $\Theta_D \equiv 2\left[\frac{R_1}{{\rm AU}}\right]\pi_p$. Sub-percent precision on the angular size is achievable, even when accounting for uncertainties in the limb-darkening coefficients \citep{vonBraun2014}, but only for relatively bright (typically, $V \lesssim 10$) stars.

\subsubsection{\label{subsec:seismoscale}  Asteroseismology and Granulation}
Asteroseismology provides several ways to determine the stellar density and surface gravity, which can be combined to find the stellar mass and radius. For stars with low signal-to-noise ratio photometry (resulting in a low-resolution frequency spectrum), one can measure the average large frequency spacing $\langle \Delta \nu \rangle$ and the frequency of maximum oscillation power $\nu_{max}$ for oscillations driven by stellar surface convection. For solar-like stars on the main sequence, empirical scaling relations from \citet{Kjeldsen1995} map these observables to physical parameters:
\begin{equation}
\label{eq:dnu} \frac{\rho}{\rho_{\sun}} = \left(\frac{\langle \Delta \nu \rangle}{\langle\Delta \nu\rangle_{\sun}}\right)^2
\end{equation}
\begin{equation}
\label{eq:numax} \frac{g_1}{g_{\sun}} = \left(\frac{\nu_{max}}{\nu_{max,\sun}}\right)\left(\frac{\teff{1}}{T_{\rm eff,\odot}}\right)^{1/2}.
\end{equation}
We note that the above scaling relations, as written, may not be accurate for some stars: see the \citet{White2011} effective temperature correction for dwarfs and subgiants (which introduces a factor of order $\mathcal{O}(\teff{1})$ into Equation \ref{eq:dnu}), the metallicity-induced deviations from these relations found in evolved stars by \citet{Epstein2014}, and the nonlinear effective temperature and metallicity correction to the large frequency spacing reference value ($\Delta \nu_{\sun}$ in the above equations) from \citet{Guggenberger2016}. The solar reference values are not inherent to the Sun but instead differ from one pipeline to the next, so we do not scale these relations by specific numerical solar values. Typical reference values are approximately $\Delta \nu_{\sun} \sim 135 \mu$Hz and $\nu_{max,\sun} \sim 3100 \mu$Hz, per Table 1 of \citet{Pinsono2018}, who discussed this in more detail.

Nevertheless, as our main goal is simply to illustrate how one can use asteroseismic measurements to constrain the mass and radius of the binary components, we perform our analysis using the above relations for simplicity.
For fixed $\nu_{max}$ and $\teff{1}$, Equation \eqref{eq:numax} becomes a mass-radius relation where $M \propto R^2$; a density from, e.g. the asteroseismic large frequency spacing or the eclipse photometry can break the degeneracy. Analogously to Section \ref{subsec:specscale}, the primary mass and radius are given by
\begin{equation}
\label{eq:m1aphys0} \frac{M_1}{M_{\sun}} = \frac{9}{16 \pi^2 G^3}\left(\frac{g_1}{g_{\sun}}\right)^3\left(\frac{\rho_1}{\rho_{\sun}}\right)^{-2}
\end{equation}
and
\begin{equation}
\label{eq:r1aphys0} \frac{R_1}{R_{\sun}} = \frac{3}{4\pi G}\frac{g_1}{\rho_1}\frac{\rho_{\sun}}{g_{\sun}}
\end{equation}.
For solar-like dwarfs (roughly spectral types F through K), the asteroseismic scaling relations have been tested theoretically to $\sim 2\%$ and $\sim 5\%$ and empirically to $\sim 4\%$ and $\sim 10\%$ \citep{Huber2013}. More recently, \citet{Coelho2015} found these scaling relations to have a random scatter in $\nu_{max}$ of 1.5\% for effective temperatures between $\sim 5,600-7,000$ K.  To see how such a scatter affects the mass and radius predicted by these relations, we solve for these parameters in the equations above:
\begin{equation}
\label{eq:m1aphys} \frac{M_1}{M_{\sun}} = \frac{9}{16 \pi^2 G^3}\left(\frac{\nu_{max}}{\nu_{max,\sun}}\right)^3\left(\frac{\langle \Delta \nu\rangle}{\langle \Delta \nu \rangle_{\sun}}\right)^4\left(\frac{\teff{1}}{T_{\rm eff,\odot}}\right)^{3/2}
\end{equation}

\begin{equation}
\label{eq:r1aphys} \frac{R_1}{R_{\sun}} = \frac{3}{4\pi G}\left(\frac{\nu_{max}}{\nu_{max,\sun}}\right)\left(\frac{\langle \Delta \nu \rangle}{\langle \Delta \nu \rangle_{\sun}}\right)^{-2}\left(\frac{\teff{1}}{T_{\rm eff,\odot}}\right)^{1/2}.
\end{equation}

 We see that a 1.5\% uncertainty in $\nu_{max}$ is amplified by a factor of three before entering the uncertainty in the mass, either directly from the expressions above or from combining the asteroseismic radius with the density calculated from the eclipse measurements. Errors on individual measurements of $\nu_{max}$ and $\langle \Delta \nu \rangle$ can further increase the uncertainty on $M_1$ and $R_1$. However, if the target star is sufficiently bright and the observing cadence sufficiently rapid, then the signal-to-noise is increased and one can match the observed frequency spectrum to those predicted by asteroseismic stellar models; for \emph{Kepler} stars, fitting for individual frequencies (\citealt{Deheuvels2012,Metcalfe2014}) and "peak-bagging" (\citealt{Lund2017,Silva2017}) reduce the uncertainty on the mass and radius by a factor of two relative to those derived from grid-based models of the global stellar oscillation properties and by a factor of three relative to those predicted from scaling relations. This procedure permits more direct inferences of the stellar mass, radius, age, and metallicity than do the scaling relations.

In addition to seismology, the primary star's surface gravity can be determined from tracers of granulation -- either the root-mean-square variations in the eclipse light curve on timescales longer than eight hours (dubbed ``flicker''; \citealt{Bastien2016}) or from isolating the granulation and oscillation signal and calculating the timescale for its autocorrelation function to fall to zero \citep{Kallinger2016}. The latter can provide gravities to 4\% but requires short-cadence observations (such as {\it Kepler} short-cadence data) for dwarfs, which oscillate on sub-hour timescales \citep{Kallinger2016}; the former can provide \logg\ to 0.1 dex but requires only long-cadence data. 

With the primary's physical parameters determined from asteroseismology or one of the granulation tracers, the companion's parameters are easily deduced: $M_2$ comes from Equation \eqref{eq:m2rv} and $R_2 = R_1\sqrt{\delta}$.

We note here that asteroseismology alone directly gives a bulk density for the primary star that is independent of the orbital eccentricity. Thus, under our assumption that $k$ is small, any discrepancy between the asteroseismic and eclipse densities can shed light on the eccentricity of the system (\citealt{Tingley2011,Huber2013}.

\subsection{\label{sec:models}Indirect Inference Methods of Mass and Radius}
When the methods in the previous section are unavailable for a star or stellar system, one can use model-dependent methods to infer the physical properties of the star. Namely, one can use empirically calibrated relations between physical parameters  or stellar isochrones generated by evolutionary models (both subject to the inherent assumption that the star is typical of the stars used to calibrate the relations or generate the isochrones). To compare the precision achievable by the direct methods discussed in Section \ref{sec:scale} to that of empirically calibrated relations and isochrones
-- specifically, to see if the precision from the former is sufficient to improve or test the latter methods -- we examine the mass-radius relations for Sun-like stars produced by empirical relations and isochrones.

\subsubsection{\label{subsec:MLscale}Empirical Relations}
We now determine what constraints the empirical relations of \citet{Torres2010} place on the mass and radius of a star, given a measurement of the star's metallicity and effective temperature and assuming a near-solar surface gravity. For this section, we assume that our stars are typical of the stars used to calibrate the empirical relations we examine. \citet{Torres2010} analyzed 95 detached binaries with precise masses and radii and derived polynomial expressions for the masses and radii of stars (in solar units) as functions of their effective temperatures, surface gravities, and metallicities: 
\begin{equation}
\begin{aligned}
\label{eq:mtorres} \log M_1 = & a_1 + a_2X + a_3X^2 + a_4X^3 + \\
& a_5\logg^2 + a_6\logg^3 + a_7[{\rm Fe/H}]
\end{aligned}
\end{equation}
\begin{equation}
\begin{aligned}
\label{eq:rtorres} \log R_1 = & b_1 + b_2X + b_3X^2 + b_4X^3 + \\
& b_5\logg^2 + b_6\logg^3 + b_7[{\rm Fe/H}],
\end{aligned}
\end{equation}
where $X \equiv \log\teff{1} - 4.1$ and the coefficients $a_i$ and $b_i$ are listed in \citet{Torres2010}. The fundamental assumption that \citet{Torres2010} made in deriving these relations is that the mass and metallicity of a star does not change substantially as the star evolves off of the main sequence, and thus \logg\ and \teff\, can be used as a proxy for the age of the star. 

\begin{figure*}
    \centering
    \includegraphics[width=1.\textwidth]{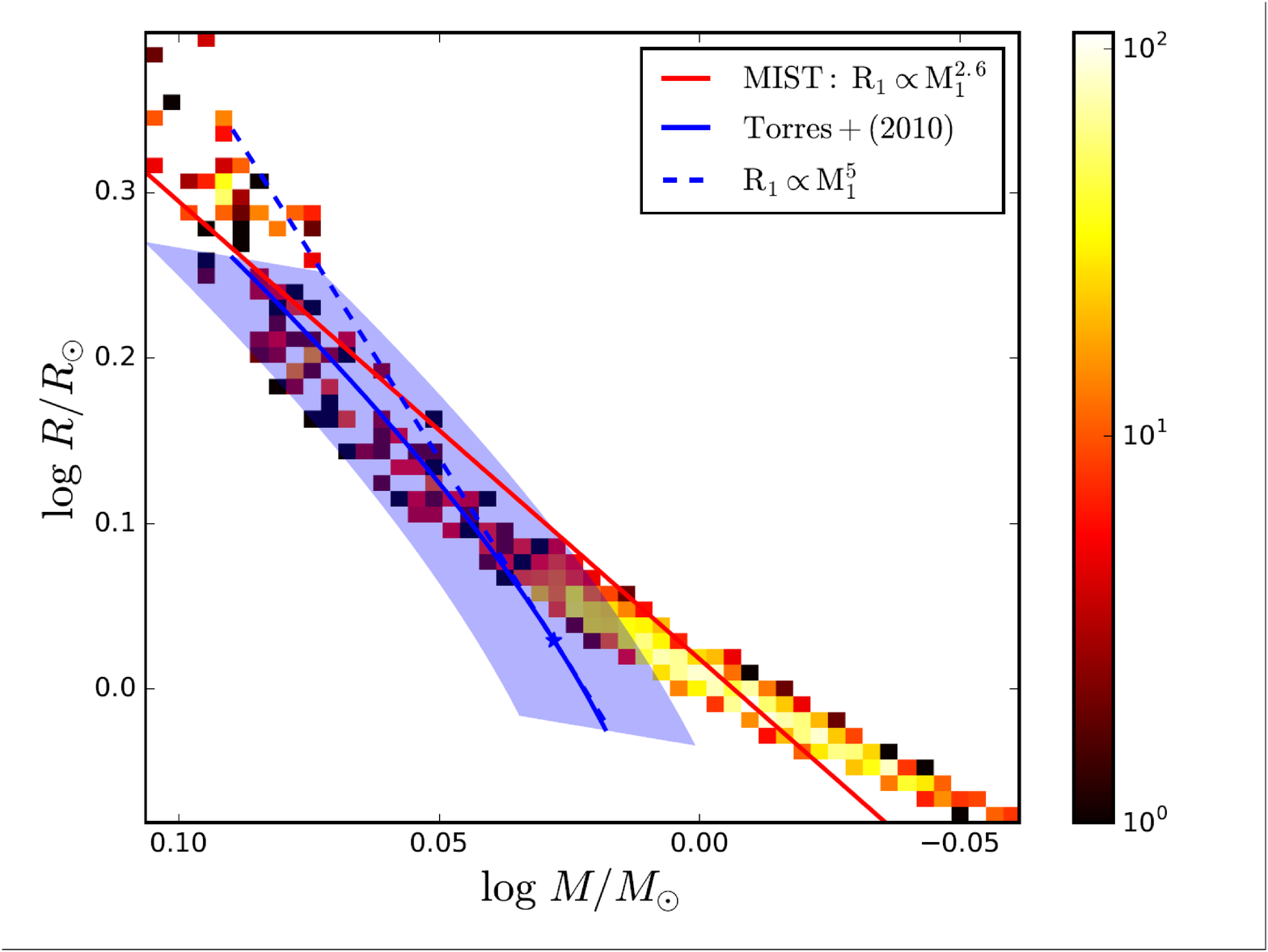}
    \caption{Heat map of radius versus mass from MIST isochrones with $\teff{1} = 5800 \pm 150$K, [Fe/H] = 0.0 $\pm$ 0.5, and solar age, with the approximate power-law fit (red line). The color bar shows the density of points in the MCMC posterior distribution per mass-radius bin. The \citet{Torres2010} mass-radius relation around solar $\logg$ (blue curve and contours) for $\teff{1} = 5800 \pm 150$K is superimposed for comparison, as is the power-law fit to the Torres relations around solar $\log g$ (dashed blue line).}
    \label{fig:iso}
\end{figure*}

To reduce the degrees of freedom in these relations and to obtain a quantity whose role is analogous to $\logg$ from the preceding subsections, we perturb Equations \eqref{eq:mtorres} and \eqref{eq:rtorres} around $\logg = \log{g_{\sun}}$ to obtain
\begin{equation}
\label{eq:mtorresp} \log M_1 \approx A + B(\logg - \log(g_{\sun}))
\end{equation}
and
\begin{equation}
\label{eq:rottresp} \log R_1 \approx C + D(\logg - \log(g_{\sun})),
\end{equation}
where $A \equiv a_1 + a_2X + a_3X^2 + a_4X^3 + a_5\log(g_{\sun})^2 + a_6\log(g_{\sun})^3 + a_7[{\rm Fe/H}]$, $B \equiv 2a_5\log(g_{\sun}) + 3a_6\log(g_{\sun})^2$, $C \equiv b_1 + b_2X + b_3X^2 + b_4X^3 + b_5\log(g_{\sun})^2 + b_6\log(g_{\sun})^3 + b_7[{\rm Fe/H}]$, and $D \equiv 2b_5\log(g_{\sun}) + 3b_6\log(g_{\sun})^2$. By substitution, $\log M_1 \approx (B/D)\log R_1 + A - (BC/D)$; therefore, we write
\begin{equation}
\label{eq:mtorresd} M_1 \approx ZR_1^{\alpha},
\end{equation}
where $Z \equiv 10^{A - (BC/D)}$ is a function of the primary star's metallicity and effective temperature, assuming its surface gravity is near the solar value, and $\alpha \equiv B/D$. For solar values of $\teff{1}$ and [Fe/H], we find that mass is quite insensitive to radius $M_1 \propto R_1^{0.2}$. This is not unexpected, as one of the assumptions used to derive the \citet{Torres2010} relations is that the mass of the star is constant. Figure \ref{fig:iso} shows that this power law agrees with the \citet{Torres2010} mass-radius relation near the "Torres Sun" ($\teff = 5800$K, [Fe/H] = 0, $\logg$=4.4), which is more massive and larger than the actual Sun. 

Not only does this power-law scaling imply that mass is rather insensitive to radius for near-solar values (for small perturbations about the solar surface gravity), but considering that a $\logg$ measurement carries the somewhat orthogonal constraint that $M_1 \propto R_1^2$, an eclipse density, in conjunction with both empirical relations and a $\logg$ measurement, considerably tightens the uncertainties on the primary mass and radius.

From a measurement of $Z$ and the eclipse density, we recover the primary parameters,
\begin{equation}
\label{eq:m1erphys} M_1 = \left[\left(\frac{3}{4\pi}\right)^{\alpha/3}\frac{Z}{\rho_1^{\alpha/3}}\right]^{\frac{3}{3-\alpha}}
\end{equation}
and
\begin{equation}
\label{eq:r1erphys} R_1 = \left(\frac{3}{4\pi}\right)^{\frac{1}{3-\alpha}}\left(\frac{Z}{\rho_1}\right)^{\frac{1}{3-\alpha}},
\end{equation}
where $\alpha \equiv B/D$. From here, the secondary's parameters are easy to derive: $R_2 = R_1\sqrt{\delta}$ and $M_2$ is obtained by plugging $M_1$ directly into Equation \eqref{eq:m2rv}.

These equations permit a more tractable analytic estimate of the contributions to the mass and radius uncertainties from the uncertainties on the primary stellar density $\rho_1$ and the scale factor $Z$ of our approximate mass-radius relation in the neighborhood of solar surface gravity.

While the present analysis is concerned with the precision of the empirical relations, the accuracy of the relations is of tantamount importance. For example, the \citet{Torres2010} mass and radius relations have a tight, $\sim 1\%$ scatter in radius but a $\sim 6\%$ scatter in mass, suggesting that the polynomial fit to mass, $\teff, \logg,$ and $\feh$ does not sufficiently capture the relationship between these parameters. \citet{Torres2010} argued that more \feh\ determinations are needed to fit a more reliable, higher-order polynomial to these parameters: only 20 of their 95 systems have \feh\ determinations, four of which are indirect (host cluster or galaxy metallicity) and none of which are M dwarfs. Additionally, the \citet{Mann2015} radius-\teff\ relation produces radii for low-mass stars that are accurate to $\sim 10\%$, which is larger than their determined $\sim 5\%$ offset between model radii and radii from interferometry.

Furthermore, as discussed in \citet{Mann2015}, the interferometric measurements for single stars do not provide model-independent masses; in this case, deriving an M-dwarf mass-radius relation for this sample requires the inference of masses from stellar models or another empirical relation, such as the \citet{Delfosse2000} mass-luminosity relation (specifically, a relation between mass and absolute $K_S$-band magnitude). The \citet{Delfosse2000} calibration sample includes short-period low-mass EBs such as YY Geminorum and CM Draconis, so using this mass-luminosity relation requires the implicit assumption that the M dwarfs under consideration are typical of those in the calibration set, and thus that any physical processes responsible for the observation-vs-model radius and \teff\ discrepancies in the short-period EBs are present at the same level in the M dwarfs to which this relation is applied. \citet{Boyajian2012b} found evidence of a substantial offset in radius-\teff\ space between single M dwarfs and M dwarfs in binaries, but more work is needed to determine if this offset is due to heterogeneous means of determining radius and \teff\ for single and binary stars or if it indeed points to an intrinsic difference between these populations.

With sufficiently precise measurements given as inputs to empirical relations, it is possible to infer a formal uncertainty (precision) due to errors in the input parameters that is smaller than the scatter in the relations themselves, leading one to think that the inferred parameters are more reliable than they really are (when in reality, the quality of the inferred parameters is no better than the scatter). This is particularly pertinent for analyses of single-lined EBs, wherein one may consider circumventing the difficulty of inferring accurate M-dwarf parameters by applying empirical relations (e.g. the Torres relations) to the host star and then calculating the M-dwarf parameters from the inferred host-star parameters and the transit and RV data. Any inaccuracies in the host-star parameters thus propagate to the companion parameters and must be taken into account and quantified.

As such, precise, model-independent parameter measurements of low-mass EBs -- and particularly long-period systems -- are still necessary for improving the accuracy of stellar models and the fidelity of empirical relations for these stars.

\subsubsection{Isochrones}\label{subsec:isoscale}
In order to determine the constraints that isochrones place on mass and radius and to estimate how uncertainties in the inputs to the iscohrones affect these parameters, we need to derive a mass-radius relation from isochrones. To this end, we use the MESA \citep{Paxton2011} Isochrones and Stellar Tracks (MIST; \citealt{Choi2016}) via the Python \texttt{isochrones} package \citep{Morton2015}. We set priors on $\teff{1} = 5800 \pm 150$ K and [Fe/H] = $0.0 \pm 0.5$, use the MCMC routine \texttt{emcee} \citep{Foreman-Mackey2013} to sample these isochrones, and retain those samples that satisfy $\log({\rm age/[yr]}) = 9.7 \pm 0.4$, i.e. isochrones around the age of the Sun.   Figure \ref{fig:iso} shows the mass and radius posterior distributions as a heat map. We then perform a least-squares fit to the data, adopting a power-law function $M_1 = ZR_1^{\alpha}$, where $M_1$ and $R_1$ are in solar units, in analogy to the power-law function in Section \ref{subsec:MLscale}. We find that $Z\approx 0.98$ and $\alpha \approx 0.36$, which is a slightly steeper dependence of mass on radius than we found with the \citet{Torres2010} relations in Section \ref{subsec:MLscale}. This is not surprising: the isochrones account for the age and thus evolutionary state of the star, and so they also account for the change in the structure and thus radius of the star as it evolves (e.g., as the core helium fraction increases).  

With this functional form, the primary mass and radius are given by Equations \ref{eq:m1erphys} and \ref{eq:r1erphys}.

\section{Results}\label{sec:results}

\subsection{Analytic Estimates of Observable Precision}\label{sec:precision}
To estimate the contribution to the fractional uncertainty on the primary density from the fractional uncertainties on the eclipse observables, we must account for covariances between $\delta$, T, and $\tau$. Following \citet{Carter2008}, we can write the covariance matrix between these three elements as:

$\Sigma \equiv Cov(\{\delta,T,\tau\},\{\delta,T,\tau\}) =$
\begin{equation}
\label{eq:cov1} 
\frac{1}{Q^2}\left[\begin{array}{ccc}
\frac{\delta^2}{1-\theta} & -\frac{\theta}{1-\theta}\delta T & \frac{\theta}{1-\theta}\delta T \\
-\frac{\theta}{1-\theta}\delta T & \frac{\theta(2-\theta)}{1-\theta}T^2 & -\frac{\theta^2}{1-\theta}T^2 \\
\frac{\theta}{1-\theta}\delta T & -\frac{\theta^2}{1-\theta}T^2 & \frac{\theta(6-5\theta)}{1-\theta}T^2 \end{array} \right],
\end{equation}
where $Q \equiv \sqrt{\Gamma T}\delta/\sigma$ and $\theta \equiv \tau/T$. Here, $\Gamma$ is the data sampling rate (assumed to be uniform) and $\sigma$ is the standard deviation of the measurement error distribution (assumed to be Gaussian). For small $\theta$ -- which we shall assume -- it is easy to show that the fractional uncertainty on $\tau$ dominates all other fractional uncertainties and covariances; $Q^2(\sig{\tau})^2 \approx 6/\theta$, while all other quantities are of higher order in $\theta$. This will simplify the upcoming uncertainty estimates.

We also note that $Q^2(\sig{\tau_0})^2 \approx (Q^2/4)(\sig{\tau})^2$ and $Q^2(\sig{\tau_0^2})^2 = 4Q^2 (\sig{\tau_0})^2 \approx 6/\theta$, which disagrees with the expression for the variance on $\tau_0$ given in Table 3 of Carter et al. (2008).

We can then estimate the fractional precision, $\sig{\rho_1}$, as $(\sig{\rho_1})^2 \approx$ $\mathbf{J \Sigma J^T}$. Here \textbf{J} is the vector consisting of the partial derivatives of $\rho_1$ with respect to $\delta$, $T$, and $\tau$; i.e. $\mathbf{J}$  $= \left(\frac{\del \rho_1}{\del \delta},\frac{\del \rho_1}{\del T},\frac{\del \rho_1}{\del \tau}\right)$, where
\begin{equation}
\label{eq:rhoderiv} \mathbf{J} = \left(\frac{3\rho_1}{4\delta},-\frac{3\rho_1}{2T},-\frac{3\rho_1}{2\tau}\right). 
\end{equation}
Carrying out the matrix multiplication and keeping only the lowest-order term in $\theta$ yields
\begin{equation}
\label{eq:sigrho} \Sig{\rho_1} \approx \left(\frac{3}{2}\right)\Sig{\tau} \approx \left(\frac{27}{2Q^2\theta}\right)^{1/2},
\end{equation}

which is a factor of two larger than the corresponding expression given in Table 2 of \citet{Carter2008}.

These analytic error analyses do not account for the curvature of the stellar limb. They are intended to provide the reader with order-of-magnitude ``rules of thumb'' regarding which quantities contribute, and in roughly what proportion, to the achievable precision on the stellar/planetary parameter measurements. If the stellar disk has significant curvature, then the transit shape will deviate from the trapezoidal shape of the piecewise-linear model and thus affect the uncertainties on the transit observables; in such a case, using equations based on the piecewise-linear model would result in inaccuracies that we do not consider.

In such a case, the Mandel-Agol transit model would be more accurate, as it does account for the geometry of the overlap region between the two stellar disks, assuming the two bodies can be treated as circular disks (i.e. no tidal deformities); deviations from circular due to tidal effects or rotation could affect the ingress/egress duration, depending on the alignment of the system. Table \ref{tab:paramcomp} compares our analytic precision estimates to the statistical uncertainties from fitting a Mandel-Agol transit model to synthetic data for a transiting hot Jupiter; the fractional precisions are roughly comparable.

\subsubsection{\label{sec:specsig}Photometric $\rho_1$ and Spectroscopic $\log(g_1)$}
We first note that the spectroscopic surface gravity and RV semi-amplitude measurements are independent of each other, as well as of the photometric density and eclipse depth. Performing the first-order propagation of uncertainty analysis on Equations \eqref{eq:m1g} through \eqref{eq:r2g} and keeping only the lowest-order term in $\theta$ yields
\begin{equation}
\label{eq:mgfull} \Sig{M_1}^2 \approx 9\Sig{g_1}^2 +9\Sig{\tau}^2,
\end{equation}
\begin{equation}
\label{eq:rgfull} \Sig{R_1}^2 \approx \Sig{g_1}^2 + \frac{9}{4}\Sig{\tau}^2,
\end{equation}
\begin{equation}
\label{eq:r2gfull} \Sig{R_2}^2 \approx \Sig{g_1}^2 + \frac{9}{4}\Sig{\tau}^2,
\end{equation}
and
\begin{equation}
\label{eq:m2gfull} \Sig{M_2}^2 \approx 4\Sig{g_1}^2 + \Sig{K_1}^2 + 4\Sig{\tau}^2 + \Sig{\sin i}^2.
\end{equation}

We note that the uncertainty on the ingress/egress time dominates any effect that the uncertainty on the depth would contribute to the precision on $R_2$, so we forthwith ignore the depth uncertainty. From here on, we assess the impact of inclination uncertainties on $M_2$. It can be shown that $\sig{\sin i} \approx \cot^2 i (\sig{\cos i})$. Since $\cos i = 2\pi\tau_0b/P$ for a circular orbit,
\begin{equation}
\Sig{\cos i} \approx \frac{1}{2}\Sig{\tau},
\end{equation}
so
\begin{equation}
\label{eq:sinisig} \Sig{\sin i}\approx \frac{\cot^2 i}{2} \Sig{\tau}
\end{equation}

Thus, the contribution from the uncertainty on $\sin i$ becomes comparable to the contribution from the uncertainty on $\rho_1$ when $\cot^2 i \approx 4$, which occurs when $|i| \lesssim 26.57$ degrees. In this case, for a transiting system with $i \approx 90$ degrees, the uncertainty on $\sin i$ is negligible and thus contributes negligibly to the uncertainties on the transit observables. However, we note that, for a high-inclination orbit causing near-grazing or grazing eclipses, the uncertainty on the inclination can significantly affect both the precision of the transit observables as well as the accuracy of the piecewise-linear light-curve model assumed for this analysis.

\begin{figure*}
\includegraphics[width=1.\textwidth]{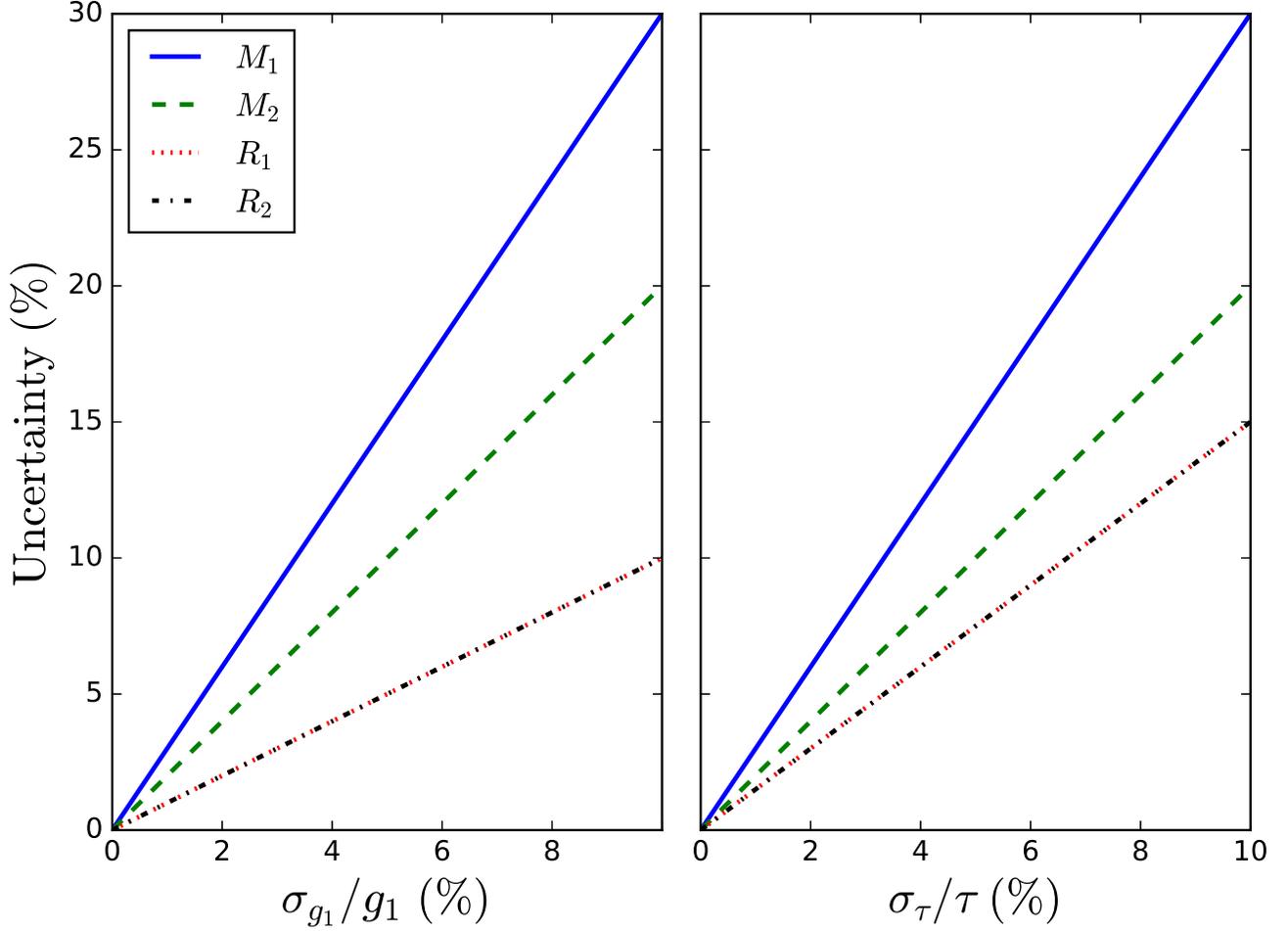}
\caption{\label{fig:specsig} 
Best achievable precision for $M_1$ (blue solid), $M_2$ (green dashed), $R_1$ (red dotted) and $R_2$ (black dot-dashed) as functions of the fractional uncertainty on surface gravity (left) and eclipse ingress/egress duration (right). Since $R_2 \propto R_1$ (related via the transit depth), any parameter that contributes to the uncertainty on $R_1$ contributes to the uncertainty on $R_2$ by the same amount.}
\end{figure*}

Figure \ref{fig:specsig} shows the allowable uncertainties needed to achieve $3\%$ uncertainty on the masses and radii. The radii are less sensitive to the uncertainty on surface gravity and ingress/egress time, and they are (as expected) independent of the uncertainty on the RV semi-amplitude. The companion mass is less sensitive to uncertainties in the these observables, but unlike $M_1$, $M_2$ obviously also depends on the primary star's RV semi-amplitude $K_1$. Also note that the $R_1$ and $R_2$ lines overlap completely: $R_2 \propto R_1$, since $R_2$ is determined from $R_1$ and the transit depth, so any parameter that contributes to the uncertainty on $R_1$ contributes to the uncertainty on $R_2$ by the same amount. 

If we impose a uniform fractional uncertainty on all four quantities (e.g. $\leq 3\%$ precision), then we see immediately that the uncertainty in $M_1$ places the tightest constraints on the requisite precision of the observables; these constraints require that $g_1$ be known to better than 1\%, or equivalently, that $\logg$ be known to better than 0.0043 in dex, since $\sigma_{\log g} \approx \sigma_g/(g\ln(10))$. Spectroscopic $\log g$ uncertainties are typically an order of magnitude larger, such as the $\sim 0.03$ dex uncertainties achieved for F, G, and K stars by \citet{Brewer2016}.

\subsubsection{\label{subsec:parallaxsig}Photometric Density and Parallax}
First, we note that the parallax measurement is independent of all other parameters, and we assume that the bolometric flux and effective temperature are as well\footnote{See \citet{StassunGaiaPlanets:2016} and \citet{StassunGaiaEB:2016} for a discussion of the minor correlations among these parameters.}. Hence, the only covariances are again between $T, \tau,$ and $\delta$, and these are small. As before, we assume that the parallax angle in radians is sufficiently small such that $\sin \pi_p \approx \pi_p$. Proceeding as before, we take Equations \eqref{eq:r1pphys}, \eqref{eq:r2pphys}, \eqref{eq:m1p}, and \eqref{eq:m2p} and obtain the following expressions for the fractional uncertainties:
\begin{equation}
\begin{aligned}
\label{eq:m1pfull} \Sig{M_1}^2 \approx\ & 36\Sig{\teff{1}}^2 + 9\Sig{\pi_p}^2 + \frac{9}{4}\Sig{\fbol}^2\\
& + \frac{9}{4}\Sig{\tau}^2,
\end{aligned}
\end{equation}
\begin{equation}
\label{eq:r1pfull} \Sig{R_1}^2 \approx 4\Sig{\teff{1}}^2 + \Sig{\pi_p}^2 + \frac{1}{4}\Sig{\fbol}^2,
\end{equation}
\begin{equation}
\begin{aligned}
\label{eq:m2pfull} \Sig{M_2}^2 \approx\ & 16\Sig{\teff{1}}^2 + 4\Sig{\pi_p}^2 + \Sig{\fbol}^2 \\
& + \Sig{K_1}^2 + \Sig{\tau}^2,
\end{aligned}
\end{equation}
and
\begin{equation}
\begin{aligned}
\label{eq:r2pfull} \Sig{R_2}^2 \approx\ & 4\Sig{\teff{1}}^2 + \Sig{\pi_p}^2 + \frac{1}{4}\Sig{\fbol}^2 \\
& + \frac{1}{4}\Sig{\delta}^2.
\end{aligned}
\end{equation}

\begin{figure*}
\includegraphics[width=1.\textwidth]{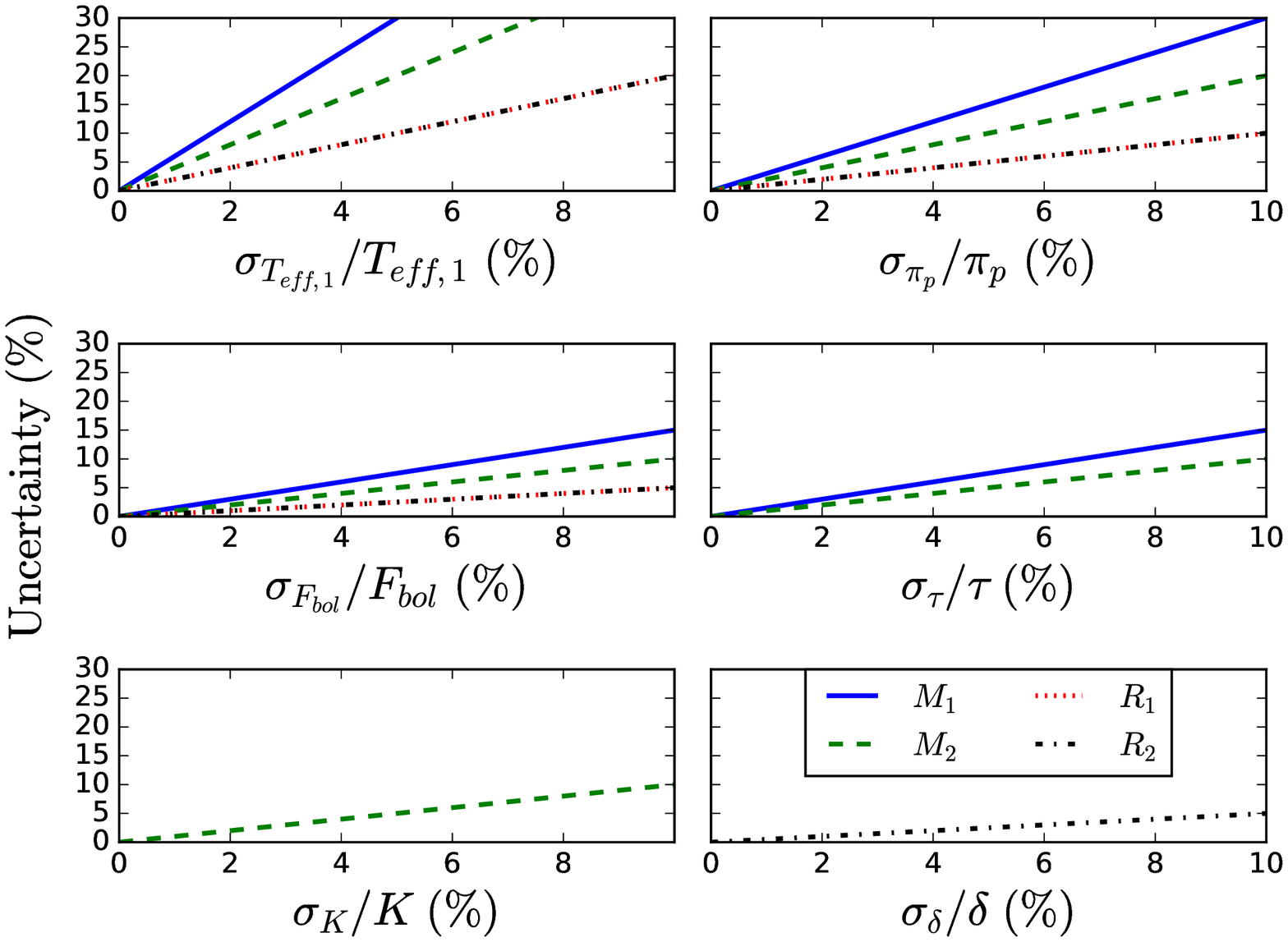}
\caption{\label{fig:parsig} 
Best achievable precision for $M_1$ (blue solid), $M_2$ (green dashed), $R_1$ (red dotted) and $R_2$ (black dot-dashed) as functions of the fractional uncertainty on the primary's effective temperature (top left), parallax angle (top right), bolometric flux (middle left), ingress/egress duration (middle right), RV semi-amplitude (bottom left), and eclipse depth (bottom right). In all but the bottom right panels, the $R_1$ and $R_2$ lines overlap: since $R_2 \propto R_1$ (related via the transit depth), any parameter that contributes to the uncertainty on $R_1$ contributes to the uncertainty on $R_2$ by the same amount.}
\end{figure*}

 Figure \ref{fig:parsig} illustrates how the precision of the masses and radii scales with the uncertainty on the observables.  In this case, the biggest contributors to the error budgets are not from the photometry or RV: rather, uncertainties on the primary's effective temperature dominate, and uncertainties on the parallax angle also contribute substantially. Indeed, one would need to know $\teff{1}$ to better than half a percent to hope to know the primary's mass to within $3\%$; even though the companion mass is proportional to $M_1^{2/3}$ (so uncertainties on $\teff{1}$ contribute less to $M_2$ than to $M_1$ by a factor of 2/3), $\teff{1}$ still needs to be known to better than 0.75\% for 3\% precision on $M_2$.
 
 Such precise effective temperatures are elusive, although some groups have demonstrated effective temperatures to relative precisions of 25 K \citep{Brewer2016}. We stress that, at such high precision, the accuracy of the \teff\ measurement and/or the underlying \teff\ scale is critical.  As an example, \citet{Casagrande2014} discussed zero-point offsets at the tens of K level between different \teff\ scales and the difficulty in determining a ``true'' \teff\ scale. For a well-calibrated, internally consistent \teff\ scale whose values differ from another \teff\ scale by some fixed amount (a zero-point offset), comparative analyses of different stars using the same \teff\ scale may be more informative than one that uses different \teff\ determinations.

Using Equation \ref{eq:sinisig}, the contribution to the uncertainty on $M_2$ from the uncertainty on $\sin i$ becomes comparable to the contribution from the uncertainty on $\rho_1$ when $\cot^2 i \approx 2$, or when $|i| \lesssim 35.26$ degrees. Again, for a transiting system with $i \approx 90$ degrees, the contribution from $\sin i$ is small, and so we ignore it. As we cautioned in the previous subsection, we warn again that the uncertainty on the inclination can significantly affect both the precision of the observable transit quantities (e.g. ingress/egress duration) and the accuracy of the light-curve model we adopt in this paper.

\subsubsection{\label{subsec:seismosig}Asteroseismology and Granulation}
With the primary's parameters $M_1$ and $R_1$ determined  from asteroseismology, flicker, and/or the granulation timescale, we can easily express the fractional uncertainties in $M_2$ and $R_2$ as
\begin{equation}
\label{eq:m2a} \Sig{M_2}^2 \approx \Sig{K_1}^2 + \frac{4}{9}\Sig{M_1}^2
\end{equation}
and
\begin{equation}
\label{eq:r2a} \Sig{R_2}^2 \approx \Sig{R_1}^2 + \frac{1}{4}\Sig{\delta}^2.
\end{equation}

In this case, asteroseismology obviates the need for precision ingress and egress measurements -- the dominant sources of uncertainty in eclipse modeling. Alongside the precision on the RV parameters, only the eclipse signal-to-noise ratio $Q$ matters, since $\sig{\delta} \sim 1/Q$.

For frequency spectrum matching, we cannot determine analytically which errors propagate into the uncertainties on the primary's mass and radius; we can, however, do so for the scaling relations. Assuming negligible covariance between the large frequency spacings and the frequency of maximum power, first-order propagation of error on Equations \eqref{eq:m1aphys} and \eqref{eq:r1aphys} reveals that

\begin{equation}
\label{eq:m1a} \Sig{M_1}^2 \approx 16\Sig{\langle \Delta \nu \rangle}^2 + 9\Sig{\nu_{\max}}^2 + \frac{9}{4}\Sig{\teff{1}}^2
\end{equation}
and
\begin{equation}
\label{eq:r1a} \Sig{R_1}^2 \approx 4\Sig{\langle \Delta \nu \rangle}^2 + \Sig{\nu_{\max}}^2 + \frac{1}{4}\Sig{\teff{1}}^2,
\end{equation}
which translate to, through Equations \eqref{eq:m2a} and \eqref{eq:r2a},
\begin{equation}
\label{eq:m2aSR} \Sig{M_2}^2 \approx \frac{64}{9}\Sig{\langle \Delta \nu \rangle}^2 + 4\Sig{\nu_{\max}}^2 + \Sig{\teff{1}}^2 + \Sig{K_1}^2
\end{equation}
and
\begin{equation}
\label{eq:r2aSR} \Sig{R_2}^2 \approx 4\Sig{\langle \Delta \nu \rangle}^2 + \Sig{\nu_{\max}}^2 + \frac{1}{4}\Sig{\teff{1}}^2 + \frac{1}{4}\Sig{\delta}^2.
\end{equation}

\begin{figure*}
\includegraphics[width=1.\textwidth]{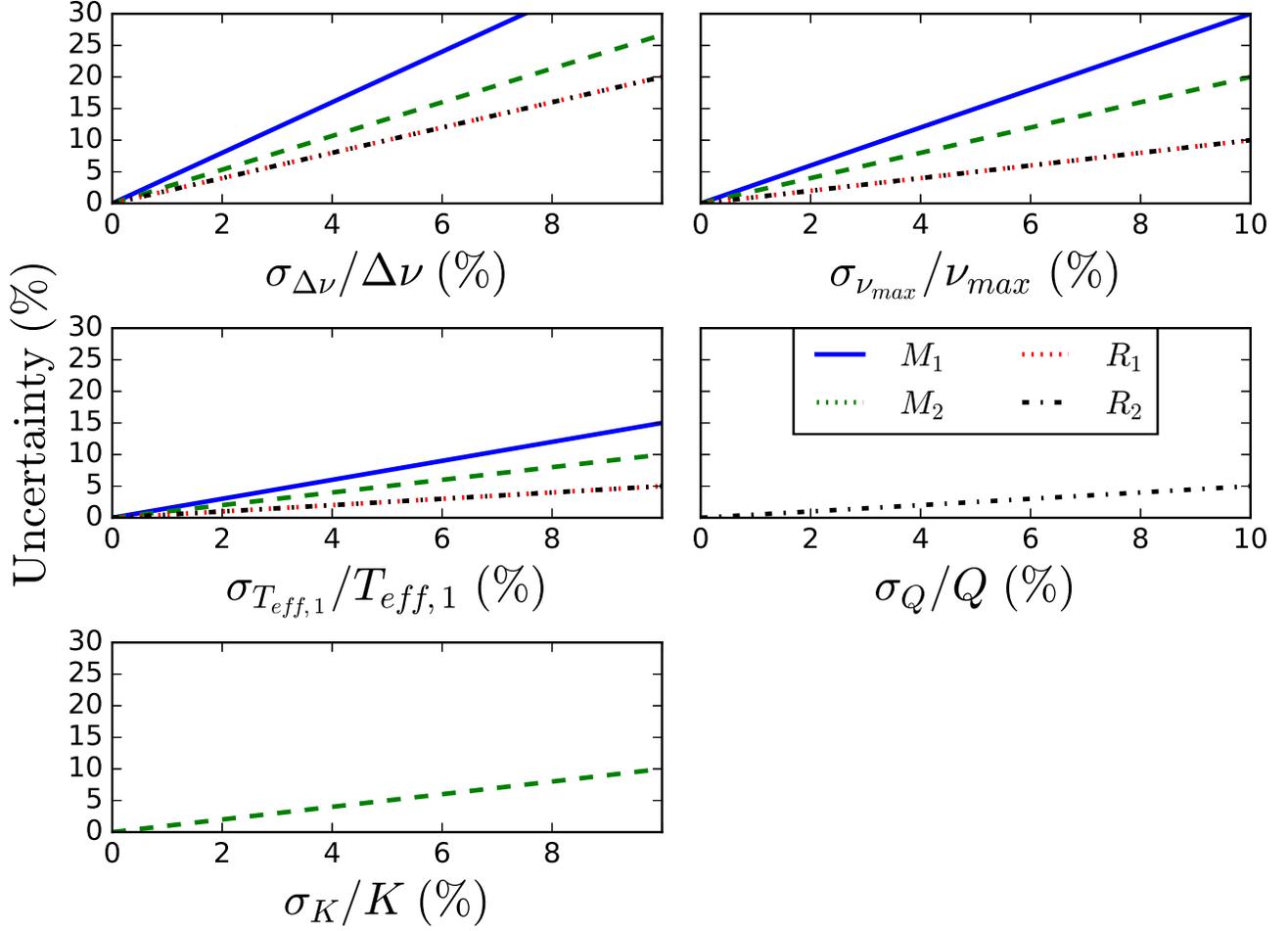}
\caption{\label{fig:asterosig} 
Best achievable precision for $M_1$ (blue solid), $M_2$ (green dashed), $R_1$ (red dotted) and $R_2$ (black dot-dashed) as functions of the fractional uncertainty on the asteroseismic large frequency spacing (top left) and peak frequency (top right) as well as on the effective temperature of the primary (bottom left) and the RV semi-amplitude and depth (bottom right). Since $R_2 \propto R_1$ (related via the transit depth), any parameter that contributes to the uncertainty on $R_1$ contributes to the uncertainty on $R_2$ by the same amount, so the $R_1$ and $R_2$ lines overlap in all but the bottom right panel (where only $R_2$ is plotted).}
\end{figure*}

The above equations illustrate clearly that the dominant sources of error on all four physical parameters are the asteroseismic quantities -- particularly the large frequency spacing, which contributes a factor of $\sim$2-4 more uncertainty than does $\nu_{max}$. This is also shown in Figure \ref{fig:asterosig}. It is typically easier to measure $\langle \Delta \nu \rangle$ than $\nu_{max}$ in noisy data \citep{Huber2013}, so the uncertainty in the former should generally be smaller than the uncertainty in the latter. With {\it Kepler} data, it is possible to measure $\Delta{\nu}$ and $\nu_{\max}$ to better than 1\%, as in the case of the Kepler-56 planetary system \citep{Huber2013Sci}.

\subsubsection{\label{subsec:MLsig}Empirical Relations}
\begin{figure*}
\includegraphics[width=1.\textwidth]{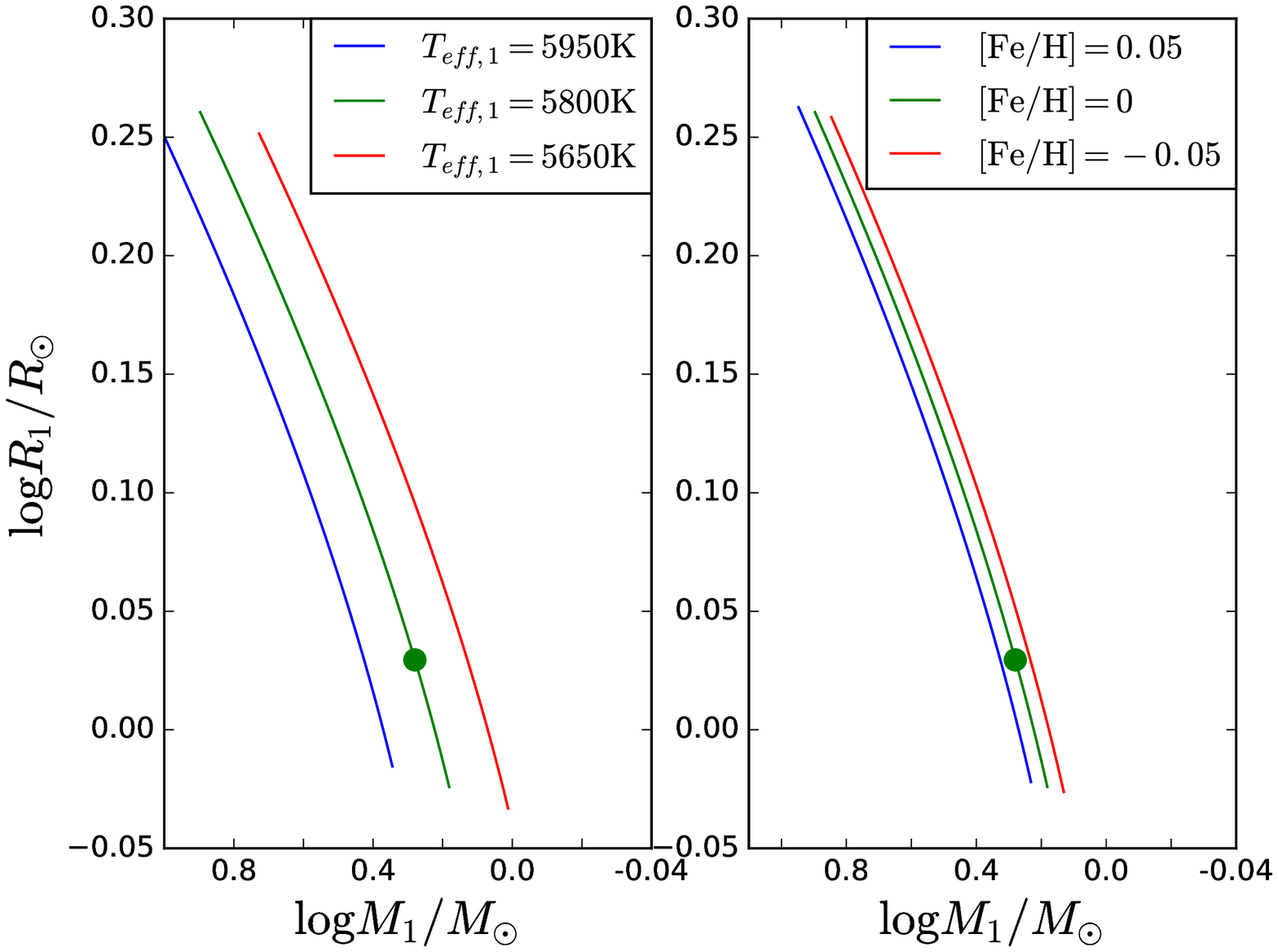}
\caption{\label{fig:ERalpha} Effect on the mass-radius relation of $\teff{1}$ (left) and [Fe/H] (right) for surface gravity near the solar value. The green circle denotes the adopted solar values of $\teff{\odot} = 5800$ K, $\log g_{\odot} = 4.4$, and [Fe/H]$_{\odot} = 0.0$.}
\end{figure*}

To make the analytic estimates more tractable, we assert that the exponent $\alpha$ in Equation \eqref{eq:mtorresd} is more or less constant for near-solar surface gravity over a small range of effective temperature and metallicity centered on solar values. Figure \ref{fig:ERalpha} shows that, for a $\sim 2.6\%$ fractional deviation in $\teff{1}$ around 5800 K and for a 0.05 dex deviation in [Fe/H] about 0, the slope of the mass-radius relation changes nearly imperceptibly. Hence, for sufficiently precise measurements of effective temperature and metallicity, we can ignore the contribution of the uncertainty in $\alpha$ to the uncertainties on the masses and radii, which will almost certainly be significantly more affected by uncertainties in $Z$, $\rho_1$, $\delta$, and $K_1$.

We first note that $Z$ is independent of the transit and RV parameters and propagate errors through Equations \eqref{eq:m1erphys} and \eqref{eq:r1erphys}:
\begin{equation}
\label{eq:m1ersig} \Sig{M_1}^2 \approx \left(\frac{3}{3-\alpha}\right)^2\Sig{Z}^2 + \left(\frac{3\alpha}{6-2\alpha}\right)^2\Sig{\tau}^2,
\end{equation}
\begin{equation}
\label{eq:r1ersig} \Sig{R_1}^2 \approx \left(\frac{3}{6-2\alpha}\right)^2\Sig{\tau}^2 + \left(\frac{1}{3-\alpha}\right)^2\Sig{Z}^2,
\end{equation}
\begin{equation}
\label{eq:m2ersig} \Sig{M_2}^2 \approx \Sig{K_1}^2 + \left(\frac{2}{3-\alpha}\right)^2\Sig{Z}^2 + \left(\frac{\alpha}{3-\alpha}\right)^2\Sig{\tau}^2,
\end{equation}
and
\begin{equation}
\label{eq:r2ersig} \Sig{R_2}^2 \approx \left(\frac{3}{6-2\alpha}\right)^2\Sig{\tau}^2 + \left(\frac{1}{3-\alpha}\right)^2\Sig{Z}^2,
\end{equation}
where we again assume that the uncertainties on the ingress/egress duration dominate the uncertainties on the primary eclipse depth. For $\alpha = 0.2$, these equations become

\begin{equation}
\label{eq:m1ersig2} \Sig{M_1}^2\approx 1.15\Sig{Z}^2 + 0.01\Sig{\tau}^2,
\end{equation}
\begin{equation}
\label{eq:r1ersig2} \Sig{R_1}^2 \approx \Sig{R_2}^2 \approx 0.29\Sig{\tau}^2 + 0.13\Sig{Z}^2,
\end{equation}
and
\begin{equation}
\label{eq:m2ersig2} \Sig{M_2}^2 \approx \Sig{K_1}^2 + 0.51\Sig{Z}^2 + 0.01\Sig{\tau}^2.
\end{equation}

\begin{figure*}
\includegraphics[width=1.\textwidth]{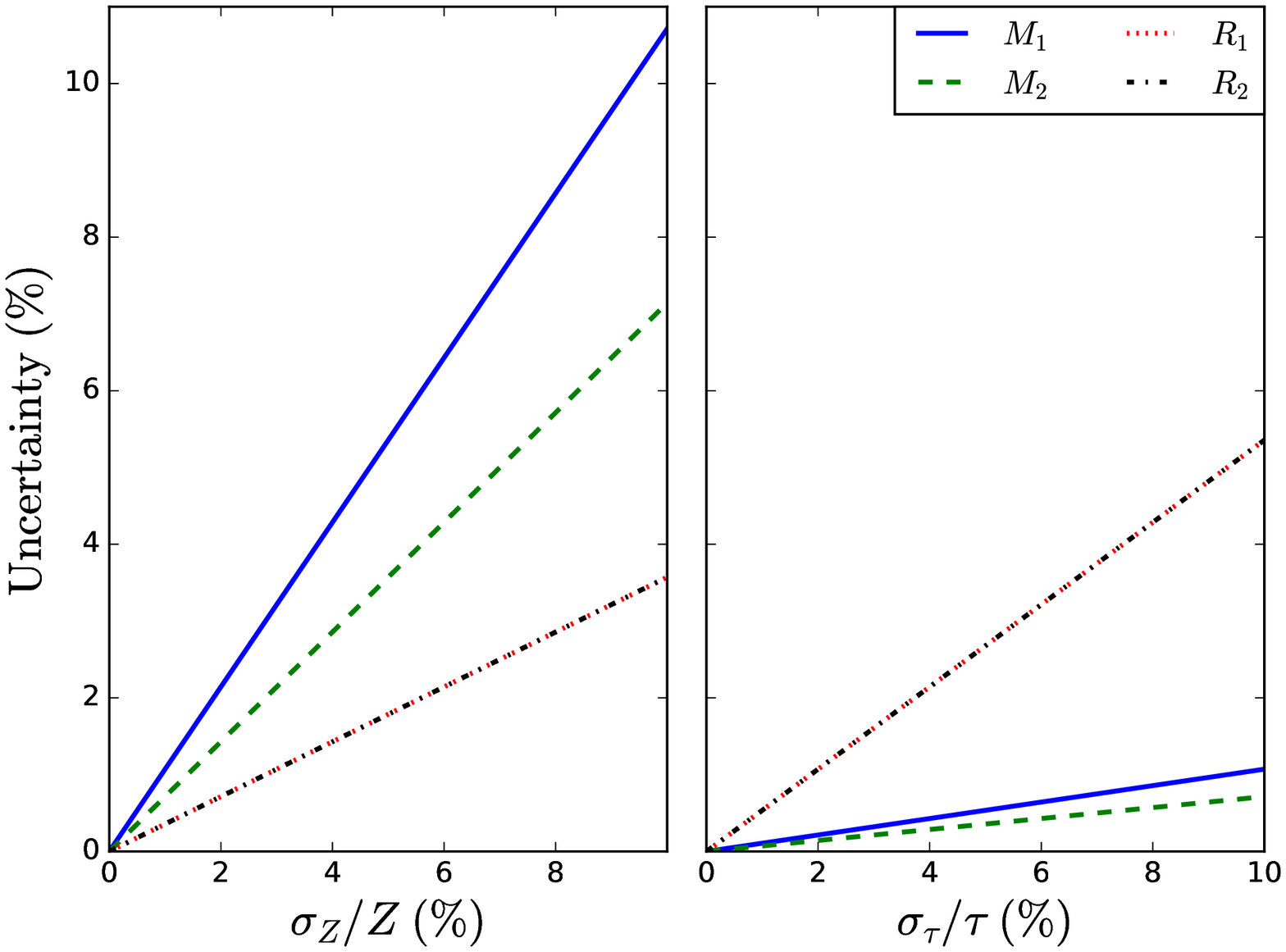}
\caption{\label{fig:ersig} Best achievable precision for $M_1$ (\emph{blue solid}), $M_2$ (\emph{green dashed}), $R_1$ (\emph{red dotted}) and $R_2$ (\emph{black dot-dashed}) as functions of the fractional uncertainty on the mass-radius relation scale factor $Z$ (\emph{left}; see Equation \ref{eq:mtorresd}) and the eclipse ingress/egress duration $\tau$ (\emph{right}).}
\end{figure*}

Thus, while the ingress/egress duration dominates the radius error budget, the uncertainty in $Z$ dominates the primary mass uncertainty, and the RV semi-amplitude error dominates the secondary mass uncertainty. Figure \ref{fig:ersig} illustrates the contributions of $\sig{Z}$ and $\sig{\tau}$ to the error budgets of the four physical parameters of interest.

To see if the requisite precision on $Z$ is achievable, we write the uncertainty in $Z$ as a function of $\teff{1}$ and [Fe/H]:
\begin{equation}
\begin{aligned}
\label{eq:Zerr1} \Sig{Z}^2 \approx & \ln^2(10) \left[ \left(3(a_4 - \frac{B}{D}b_4) \log^2 \teff{1}  \right.\right.\\
& + \left[2 (a_3 - \frac{B}{D}b_3) -24.6(a_4 - \frac{B}{D}b_4)\right]\log \teff{1} \\
& \left.\vphantom{3(a_4 - \frac{B}{D}b_4)} + (a_2 - b_2) - 8.2(a_3 -\frac{B}{D}b_3) + 50.43(a_4 - \frac{B}{D}b_4)  \right)^2 \\
& \times \Sig{\teff{1}}^2 \left.\vphantom{\left(3(a_4 - \frac{B}{D}b_4)\right) } + \left(a_7 - \frac{B}{D}b_7\right)^2\sigma_{\rm [Fe/H]}^2 \right].
\end{aligned}
\end{equation}

Since $B \approx -0.11$ and $D \approx -0.55$ for $\log g_{\sun} = 4.4$, plugging in these values and the values of the $a_i$ and $b_i$ coefficients gives

\begin{equation}
\begin{aligned}
\label{eq:Zerr} \Sig{Z}^2 \approx & \left[97.5\left(\frac{\log \teff{1}}{\log 5800 {\rm K}}\right)^2 -206\left(\frac{\log \teff{1}}{\log 5800 {\rm K}}\right) + 111\right]^2 \\ & \times \Sig{\teff{1}}^2 + 0.01\sigma_{\rm [Fe/H]}^2.
\end{aligned}
\end{equation}

\begin{figure*}
\includegraphics[width=1.\textwidth]{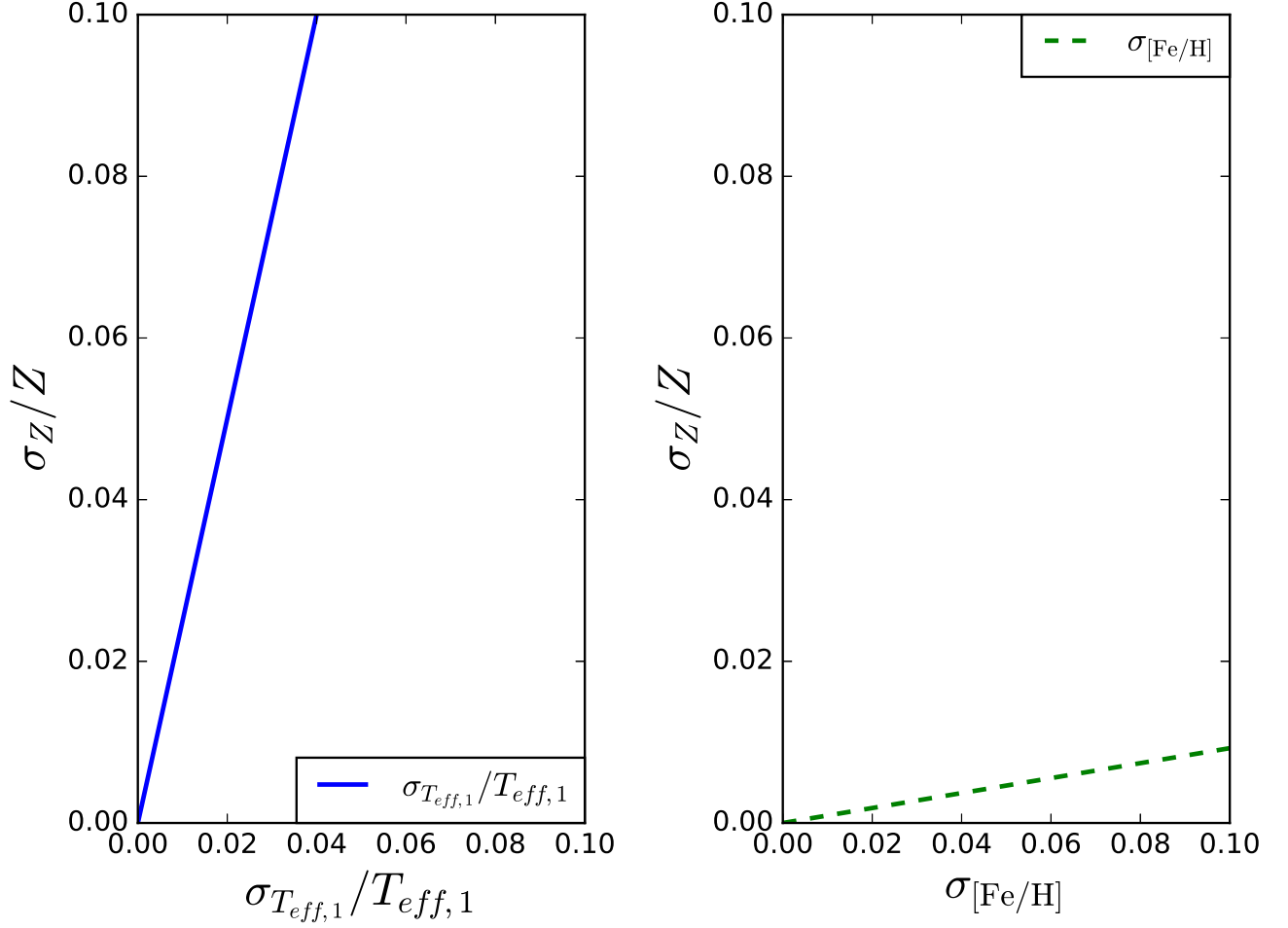}
\caption{\label{fig:erzsig} Best achievable precision for the mass-radius parameterization scale factor $Z$ (see Equation \ref{eq:mtorresd}) as a function of the fractional uncertainty in the primary star's effective temperature (blue solid line) and the uncertainty in dex in the primary star's metallicity (green dashed line).}
\end{figure*}

For $\teff{1} = 5800$ K, $\sig{\teff{1}}$ contributes to the fractional uncertainty on $Z$ by a factor of 2.5. Meanwhile, a 0.1 dex spread in metallicity yields $\sig{Z} \approx 0.01\%$. Figure \ref{fig:erzsig} illustrates these contributions for $\teff{1} = 5800$ K. Hence, $Z$ can be calculated rather precisely for Sun-like stars.

\subsubsection{Isochrones}
\label{sec:isochronesig}
\begin{figure*}
\includegraphics[width=1.\textwidth]{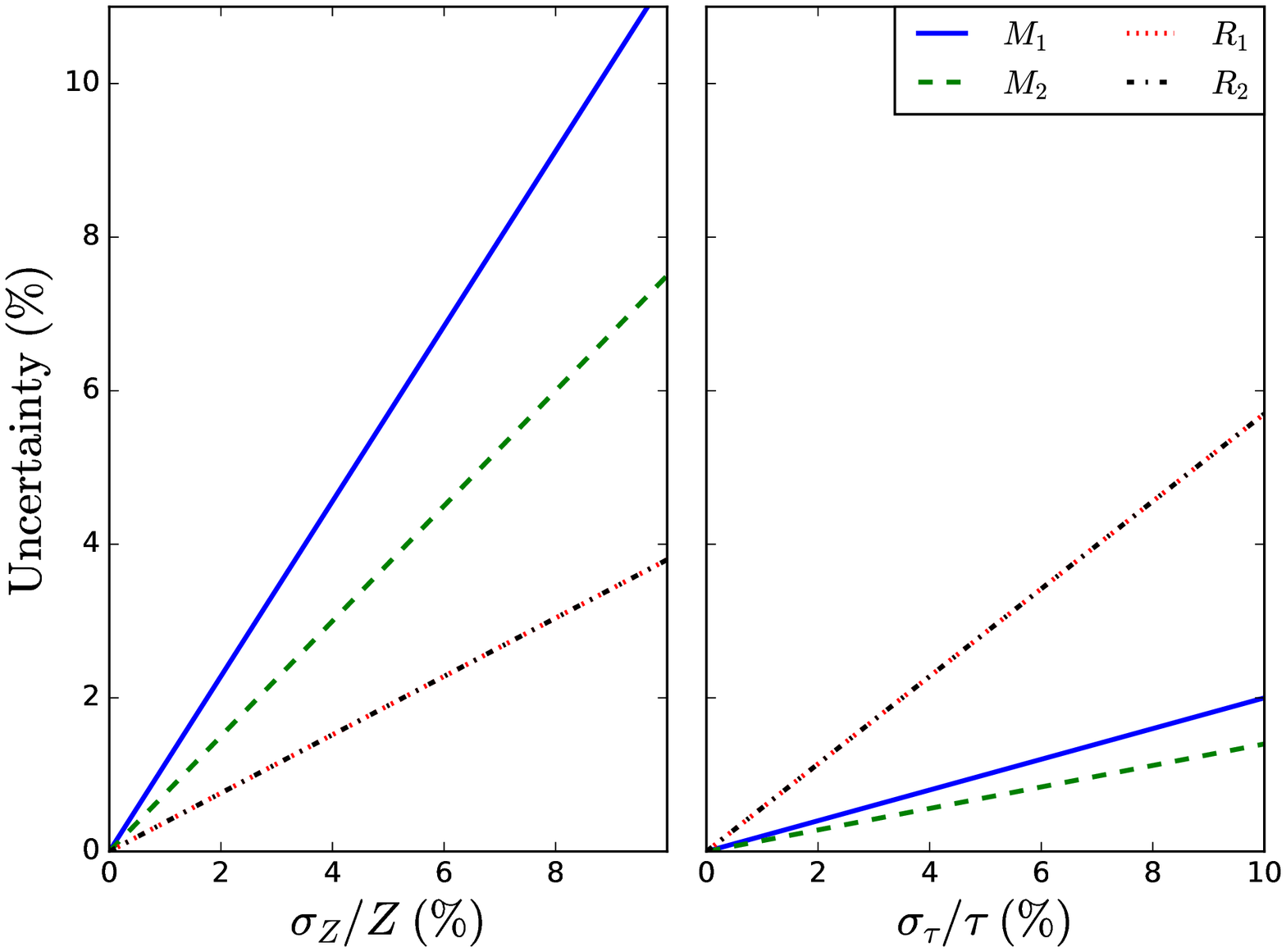}
\caption{\label{fig:isosig} Best achievable precision for $M_1$ (blue solid), $M_2$ (green dashed), $R_1$ (red dotted) and $R_2$ (black dot-dashed) as functions of the fractional uncertainty on the isochrone parameterization coefficient $Z$ (left) and the eclipse ingress/egress duration $\tau$ (right).}
\end{figure*}

From our least-squares fit in Section \ref{subsec:isoscale}, we find that a 150 K spread in effective temperature and 0.5 dex spread in metallicity produces a 2.8\% uncertainty in $Z$ and a 0.5\% uncertainty in $\alpha$; the latter implies that, as with the empirical relations in Section \ref{subsec:MLsig}, relatively precise effective temperature and metallicity measurements allow us to ignore the contribution of the uncertainty in the power-law index ($\alpha$) to the uncertainty on the masses and radii. As a result, Equations \ref{eq:m1ersig}, \ref{eq:r1ersig}, \ref{eq:m2ersig}, and \ref{eq:r2ersig} apply for the fractional uncertainties on the masses and radii. With $\alpha \approx 0.36$, these equations become
\begin{equation}
\label{eq:m1isosig} \Sig{M_1}^2\approx 1.29\Sig{Z}^2 + 0.04\Sig{\tau}^2,
\end{equation}
\begin{equation}
\label{eq:r1isosig} \Sig{R_1}^2 \approx \Sig{R_2}^2 \approx 0.32\Sig{\tau}^2 + 0.14\Sig{Z}^2,
\end{equation}
and
\begin{equation}
\label{eq:m2isosig2} \Sig{M_2}^2 \approx \Sig{K_1}^2 + 0.57\Sig{Z}^2 + 0.02\Sig{\tau}^2
\end{equation}
Figure \ref{fig:isosig} shows how uncertainties on $Z$ and the ingress/egress duration contribute to the uncertainties on the masses on radii. As with the empirical relations, $Z$ has a more significant impact on the mass uncertainties than the ingress/egress duration does, but the latter dominates the radius uncertainties.

\subsection{Example: A Transiting Hot Jupiter}\label{sec:HJ}
\begin{deluxetable*}{llccc}
\tabletypesize{\scriptsize}
\tablecaption{Median Values and 68\% Confidence Interval Limits for a Fiducial Hot Jupiter System.}
\tablehead{\colhead{Parameter} & \colhead{Units} & \multicolumn{3}{c}{Values}\\ & & $e=b=0$ & $e=0.5,b=0$ & $e=0,b=0.75$}
\startdata
\sidehead{Stellar Parameters}
~~~~$M_*$ &Mass (\msun) &$1.146^{+0.075}_{-0.092}$&$1.20^{+0.21}_{-0.23}$&$1.18^{+0.23}_{-0.19}$\\
~~~~$R_*$ &Radius (\rsun) &$1.046^{+0.017}_{-0.016}$&$1.043^{+0.017}_{-0.018}$&$1.047^{+0.018}_{-0.017}$\\
~~~~$\rho_*$ &Density (cgs) &$1.424^{+0.049}_{-0.097}$&$1.49^{+0.23}_{-0.28}$&$1.45^{+0.26}_{-0.21}$\\
~~~~$\log{g}$ &Surface gravity (cgs) &$4.461^{+0.017}_{-0.031}$&$4.480^{+0.064}_{-0.089}$&$4.469^{+0.073}_{-0.070}$\\
~~~~$T_{eff}$ &Effective temperature (K) &$5710^{+160}_{-140}$&$5740^{+160}_{-150}$&$5700^{+170}_{-140}$\\
~~~~[Fe/H] &Metallicity  &$-0.001\pm0.096$&$-0.01\pm0.10$&$0.004^{+0.097}_{-0.099}$\\
~~~~$L_*$ &Luminosity (\lsun) &$1.050^{+0.091}_{-0.10}$&$1.065^{+0.092}_{-0.10}$&$1.044^{+0.093}_{-0.10}$\\
~~~~$A_v$ &$V$-band extinction  &$0.067^{+0.070}_{-0.046}$&$0.073^{+0.071}_{-0.050}$&$0.065^{+0.069}_{-0.045}$\\
~~~~${\rm Error\ scaling}$ &Error scaling  &$1.47^{+0.57}_{-0.35}$&$1.48^{+0.60}_{-0.36}$&$1.48^{+0.60}_{-0.35}$\\
~~~~$d$ &Distance (pc) &$149.99\pm0.11$&$149.99\pm0.11$&$149.99\pm0.11$\\
~~~~$\pi$ &Parallax (mas) &$6.6670^{+0.0050}_{-0.0051}$&$6.6669^{+0.0049}_{-0.0050}$&$6.6670\pm0.0050$\\
\sidehead{Planetary Parameters}
~~~~$a$ &Semi-major axis (AU) &$0.0531^{+0.0010}_{-0.0013}$&$0.0537^{+0.0025}_{-0.0030}$&$0.0535^{+0.0028}_{-0.0026}$\\
~~~~$P$ &Period (days) &$2.980\pm0.024$&$2.966^{+0.10}_{-0.089}$&$2.968^{+0.039}_{-0.038}$\\
~~~~$M_P$ &Mass (\mj) &$1.098^{+0.050}_{-0.060}$ & $1.142^{0.135}_{-0.154}$ & $1.133^{+0.148}_{-0.126}$\\
~~~~$R_P$ &Radius (\rj) &$1.036\pm0.018$ & $1.038^{+0.022}_{-0.021}$ & $1.032^{+0.024}_{-0.025}$\\
~~~~$e$ &Eccentricity  &\nodata &$0.493\pm0.027$&\nodata\\
~~~~$\omega_*$ &Argument of periastron (deg) & \nodata &$115.6\pm2.6$&\nodata\\
~~~~$i$ &Inclination (deg) &$89.16^{+0.60}_{-0.86}$&$86.8^{+2.2}_{-2.5}$&$85.05^{+0.49}_{-0.48}$\\
~~~~$\rho_P$ &Density (cgs) &$1.314^{+0.055}_{-0.080}$&$1.36^{+0.17}_{-0.21}$&$1.36^{+0.22}_{-0.18}$\\
~~~~$logg_P$ &Surface gravity  &$4.470^{+0.014}_{-0.025}$&$4.485^{+0.049}_{-0.070}$&$4.483^{+0.059}_{-0.057}$\\
~~~~$\fave$ &Incident flux (\fluxcgs) &$0.508^{+0.061}_{-0.052}$&$0.402^{+0.065}_{-0.055}$&$0.496^{+0.078}_{-0.075}$\\
~~~~$T_C$ &Time of transit (\bjdtdb) &$-0.00005\pm0.00027$&$0.00007^{+0.00023}_{-0.00025}$&$-0.00016\pm0.00036$\\
~~~~$T_P$ &Time of periastron (\bjdtdb) &$-0.00005\pm0.00027$&$0.0639^{+0.0092}_{-0.0086}$&$-0.00016\pm0.00036$\\
~~~~$T_S$ &Time of eclipse (\bjdtdb) &$1.490\pm0.012$&$1.039^{+0.045}_{-0.040}$&$1.484^{+0.020}_{-0.019}$\\
~~~~$T_A$ &Time of ascending node (\bjdtdb) &$-0.7449^{+0.0059}_{-0.0061}$&$-0.410^{+0.025}_{-0.027}$&$-0.7422^{+0.0095}_{-0.0098}$\\
~~~~$T_D$ &Time of descending node (\bjdtdb) &$0.7448^{+0.0061}_{-0.0059}$&$0.246^{+0.016}_{-0.015}$&$0.7419^{+0.0098}_{-0.0095}$\\
~~~~$K_1$ &Stellar RV semi-amplitude (m/s) &$141.9^{+2.3}_{-2.2}$&$165.0^{+4.5}_{-4.7}$&$143.4\pm3.7$\\
~~~~$ecos{\omega_*}$ &  &\nodata&$-0.213\pm0.021$&\nodata\\
~~~~$esin{\omega_*}$ &  &\nodata&$0.444\pm0.028$&\nodata\\
~~~~$M_P\sin i$ &Minimum mass (\msun) &$0.001048^{+4.8e-05}_{-5.7e-05}$&$0.001087^{+0.000129}_{-0.000150}$&$0.001078^{+0.000141}_{-0.000120}$\\
~~~~$M_P/M_*$ &Mass ratio  &$0.000918^{+0.000030}_{-0.000025}$&$0.000915^{+0.000072}_{-0.000052}$&$0.000920\pm0.000061$\\
~~~~$R_P/R_*$ &Radius of planet in stellar radii  &$0.09970^{+0.00087}_{-0.00083}$&$0.1001^{+0.0013}_{-0.0011}$&$0.0993\pm0.0015$\\
~~~~$a/R_*$ &Semi-major axis in stellar radii  &$10.92^{+0.15}_{-0.22}$&$11.09^{+0.46}_{-0.59}$&$10.99^{+0.53}_{-0.50}$\\
~~~~$b$ &Impact parameter  &$0.16^{+0.16}_{-0.11}$&$0.33\pm0.22$&$0.947^{+0.048}_{-0.054}$\\
~~~~$\delta$ &Transit depth  &$0.00994^{+0.00017}_{-0.00016}$&$0.01003^{+0.00027}_{-0.00023}$&$0.00985\pm0.00030$\\
~~~~$T_{FWHM}$ &FWHM duration (days) &$0.08542^{+0.00088}_{-0.0017}$&$0.0484^{+0.0014}_{-0.0032}$&$0.0241^{+0.0073}_{-0.0029}$\\
~~~~$\tau$ &Ingress/egress duration (days) &$0.00878^{+0.00059}_{-0.00018}$&$0.00547^{+0.0012}_{-0.00046}$&$0.0229^{+0.0021}_{-0.0042}$\\
~~~~$T_{14}$ &Total duration (days) &$0.09419^{+0.00081}_{-0.0012}$&$0.0538^{+0.0011}_{-0.0021}$&$0.0482^{+0.0046}_{-0.0058}$\\
\sidehead{Wavelength Parameters}
~~~~$u_{1,Sloanz}$ &Linear limb-darkening coefficient  &$0.252^{+0.043}_{-0.041}$&$0.84^{+0.59}_{-0.58}$&$0.272^{+0.055}_{-0.053}$\\
~~~~$u_{2,Sloanz}$ &Quadratic limb-darkening coefficient  &$0.256^{+0.049}_{-0.048}$&$0.059^{+0.063}_{-0.059}$&$0.280^{+0.049}_{-0.051}$\\
\sidehead{Telescope Parameters}
~~~~$\gamma$ &Instrumental offset (m/s) &$2.3\pm1.6$&$0.0132^{+0.0087}_{-0.013}$&$-1.4\pm2.5$\\
~~~~$\sigma_J$ &RV jitter  &$-52^{+23}_{-14}$&$0.087^{+0.050}_{-0.087}$&$25^{+61}_{-37}$\\
\sidehead{Transit Parameters}
~~~~$\sigma^{2}$ &Added variance  &$0.000000019^{+0.000000091}_{-0.000000081}$&$0.0597^{+0.0033}_{-0.0022}$&$0.000000058^{+0.000000097}_{-0.000000085}$\\
~~~~$F_0$ &Baseline flux  &$0.999945\pm0.000074$&$0.0730^{+0.0043}_{-0.0028}$&$0.999808^{+0.000070}_{-0.000071}$
\enddata
\label{tab:fiducial}
\end{deluxetable*}

In order to verify our analytic estimates, we generate a simulated z' light curve, RV data, and broadband absolute photometric flux measurements for a Jupiter-like planet on a 3-day orbit around a Sun-like star. We use \texttt{VARTOOLS} \citep{Hartman2016} to inject a Mandel-Agol transit transit \citep{Mandel2002} into a flat light curve with a 1 mmag per-point precision and scatter drawn from a Gaussian centered at 0 mmag with 1 mmag dispersion sampled at a 100 s cadence. Note that this is typical of a single ground-based follow-up light curve from a small ($<1$ m) telescope. We can achieve better photometric precision by obtaining several follow-up light curves, as is typical for characterization of transiting systems from the ground. Additionally,  {\it TESS} itself will provide sub-mmag per-point precision for integration times of 2 minutes for stars with $I_C \lesssim 10$.

For the RVs, we assume 20 observations evenly spaced in phase with 10 m/s per-point precision and scatter drawn from a Gaussian with 10 m/s dispersion about a mean of 0 m/s. For the broad-band flux measurements, we adopt a distance of 150 pc and combine this with the absolute $V$- and $K_S$-band magnitude and colors for a G2V dwarf from \citet{Mamajek2013} to recover magnitudes in the {\it $B$, $V$, $R_C$, $I_C$, J, H, $K_s$}, and {\it WISE} $W1-W4$ bands, assuming no extinction. Furthermore, we assume a 5 $\mu$as uncertainty on the parallax, which is comparable to the best \emph{Gaia} end-of-mission parallax uncertainties for bright stars \citep{deBruijne2012}.

\begin{figure*}
    \centering
    \includegraphics[width=0.33\textwidth]{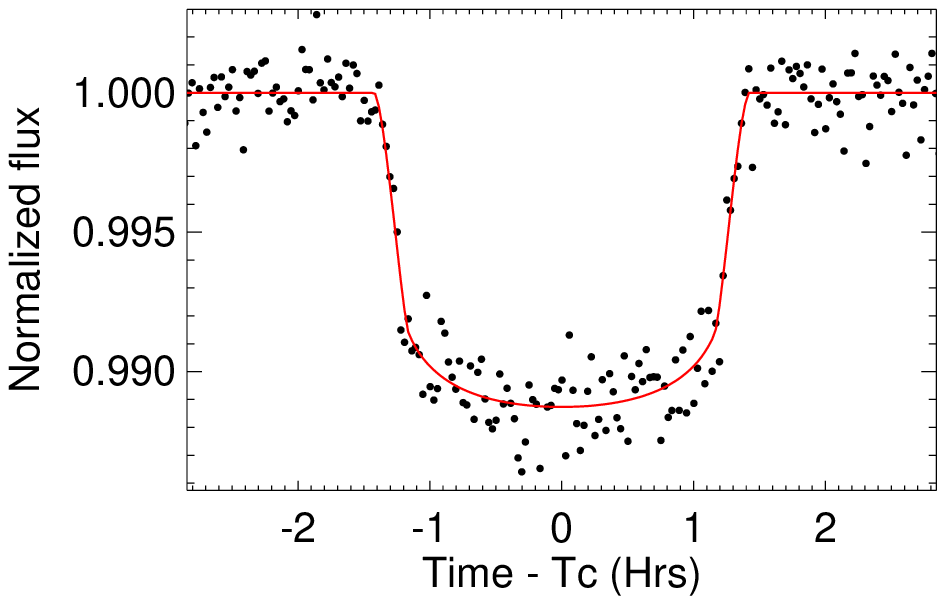}\includegraphics[width=0.33\textwidth]{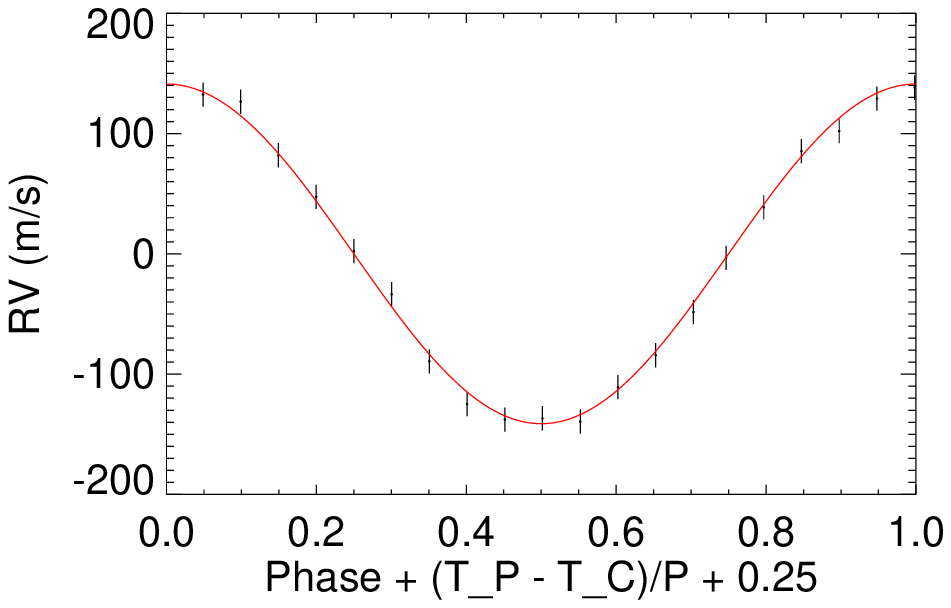}\includegraphics[width=0.33\textwidth]{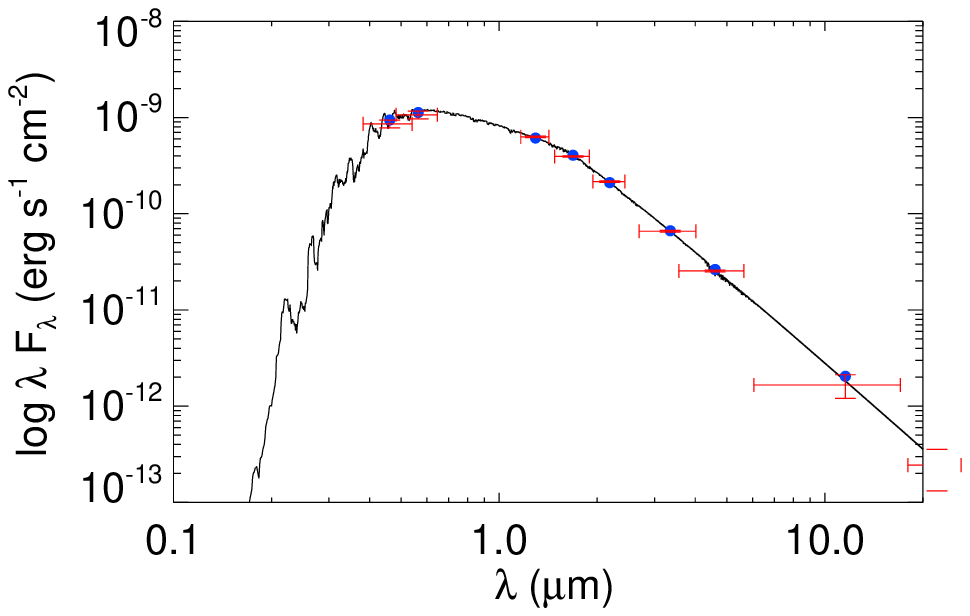}
    \includegraphics[width=0.33\textwidth]{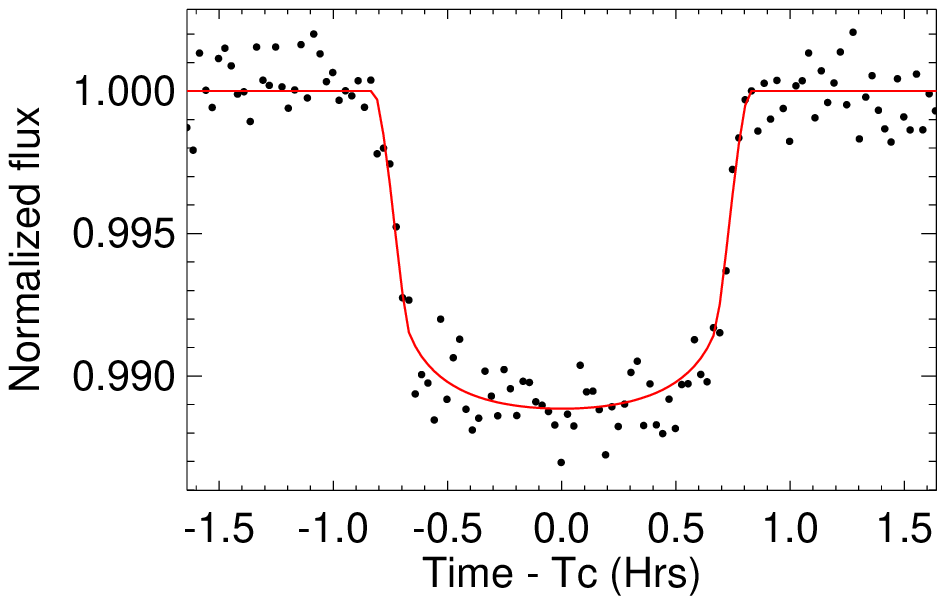}\includegraphics[width=0.33\textwidth]{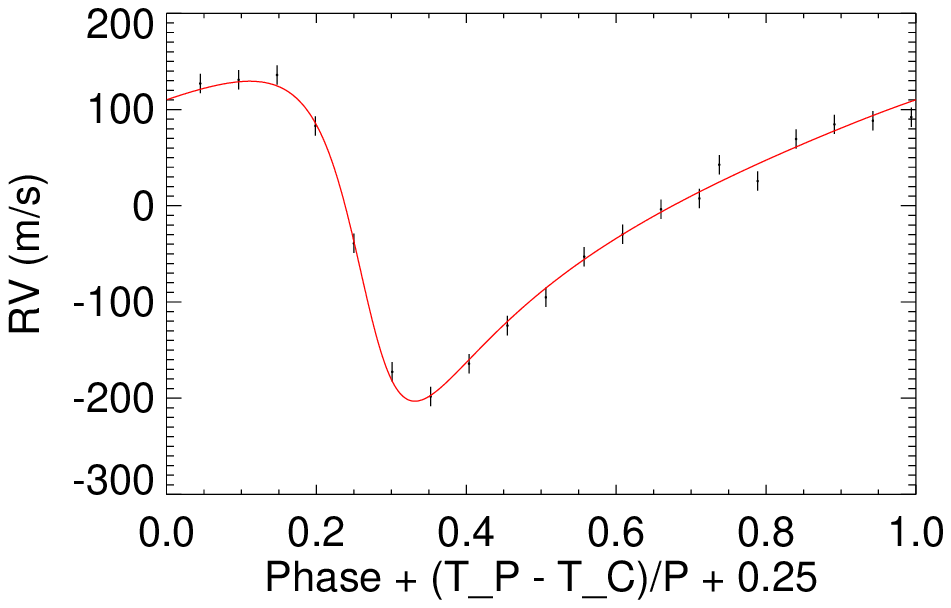}\includegraphics[width=0.33\textwidth]{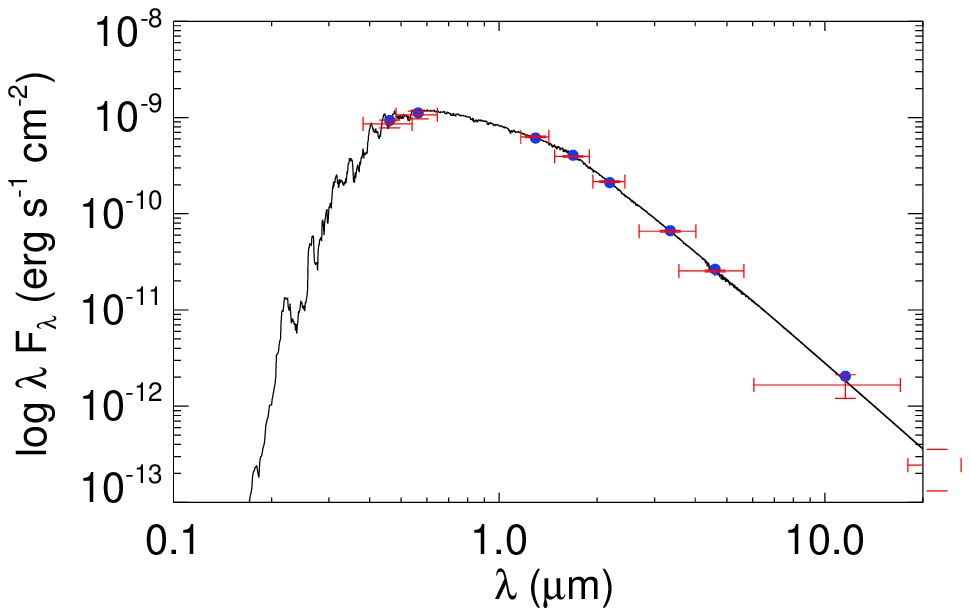}
    \includegraphics[width=0.33\textwidth]{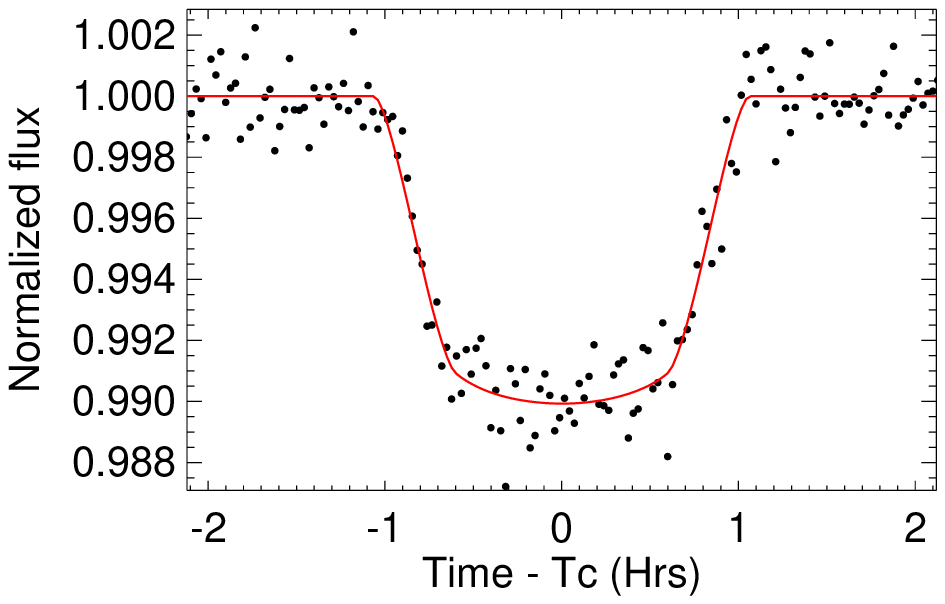}\includegraphics[width=0.33\textwidth]{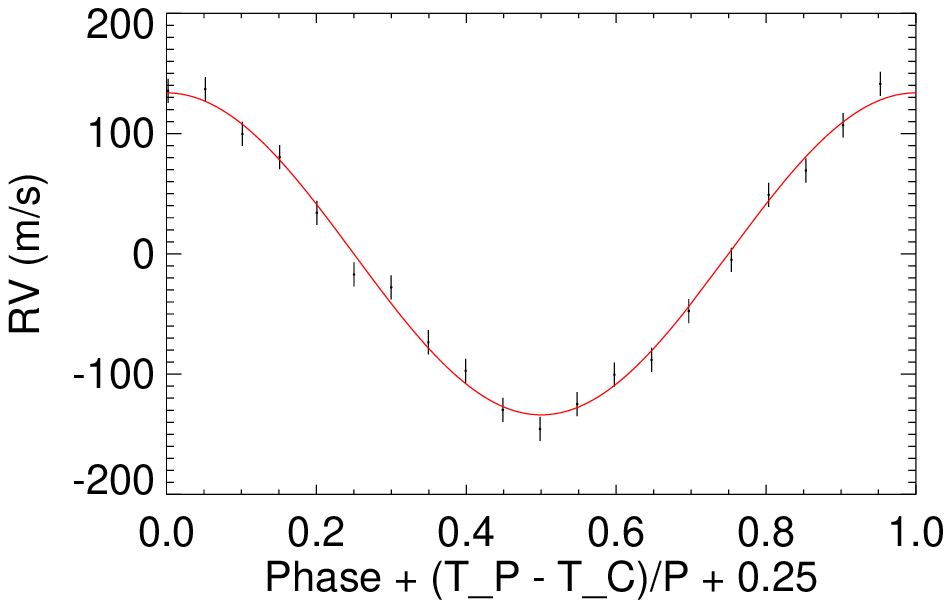}\includegraphics[width=0.33\textwidth]{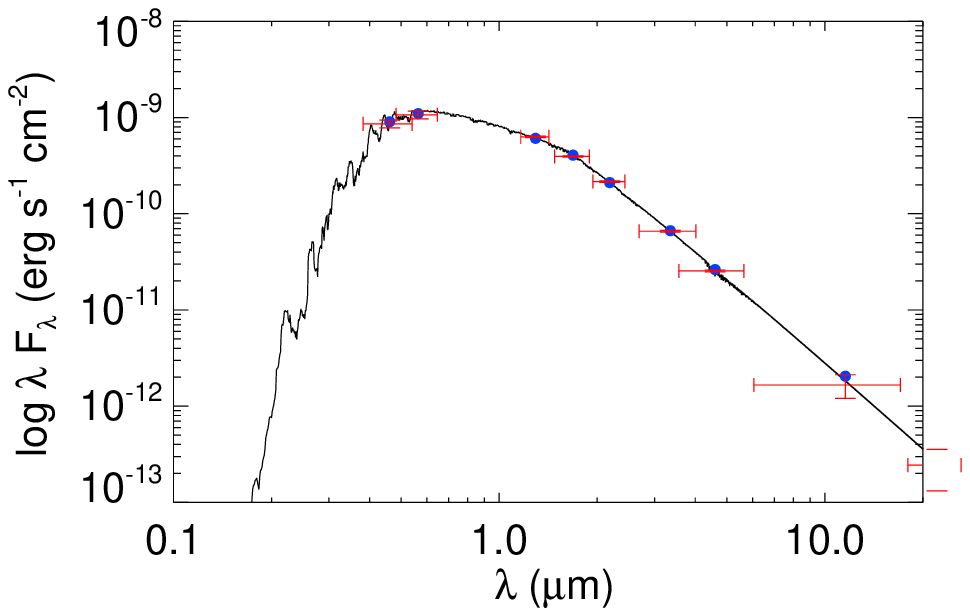}
    \caption{Transits (left), RV (middle), and SEDs (right) for a fiducial hot Jupiter transiting a Sun-like star on three different orbits: a circular, edge-on orbit (top); an eccentric, edge-on orbit (middle); and a circular orbit with high impact parameter (bottom). In all cases, the solid lines denote the model fits; in the transit and RV plots, the black dots show the mock data, while the red crosses show the mock broadband fluxes in the SED plot.}
    \label{fig:modelfits}
\end{figure*}

We examine three cases: a circular, edge-on orbit; an edge-on orbit with eccentricity $e=0.5$; and a circular orbit with impact parameter $b=0.5$. We use \texttt{EXOFASTv2} \citep{Eastman2017} to fit the RV, SED, and light curve without applying any constraints from isochrones or empirical relations. EXOFASTv2 adopts a Mandel-Agol model for the transit and uses NextGen stellar atmosphere models for the SED fitting \citep{Hauschildt1999}, interpolating the models in $\teff{1}$, $\logg$, and $\feh$. EXOFASTv2 uses a differential evolution Markov chain Monte Carlo sampling method, so the quoted uncertainties are thus the statistical uncertainties from the parameters' posterior probability distributions. Table \ref{tab:fiducial} lists the stellar and planetary parameters, while Figure \ref{fig:modelfits} shows the light curve, RV, and SED fits for each case. 

Table \ref{tab:paramcomp} compares the precision on some fit parameters to what we expect from our analytic estimates using the relations in Equation 23 of \citet{Carter2008} to estimate the errors on the observables from the signal-to-noise of the mock measurements. We find that, even with one ground-based light curve, we recover the hot Jupiter's mass and radius to better than 15\%; we expect that these uncertainties can be reduced by obtaining several high-precision ground-based light curves or by using a light curve from the {\it Kepler}, {\it K2}, or upcoming {\it TESS} survey. We further note that, despite the simplifying assumptions in our analytic estimates, these estimates agree with the numerical results to better than a factor of 2.5 in all cases and tend to be conservative -- that is, the estimated uncertainties tend to be larger than the numerical uncertainties.

Our 8\% estimate of the ingress/egress duration uncertainty for a circular orbit agrees reasonably with the 6.7\% recovered from the circular fit, but the circular fit's 6.8\% density uncertainty is markedly lower than our 13\% analytic estimate (and as such, the circular fit's mass uncertainties are smaller than expected). In fact, $(\sig{\tau}) \approx (\sig{\rho_1})$. This is likely a result of the non-negligible limb darkening; see Section \ref{sec:ld} for more discussion.

The primary and secondary radius uncertainties recovered from the fits are likely smaller than expected because of covariances in the SED parameters that our analytic approach does not take into account. A hotter star produces a larger bolometric flux, so $\teff{1}$ and $\fbol$ are positively correlated; with $R_1 \propto \sqrt{\fbol}/\teff{1}^2$, $(\sig{R_1})^2 \approx 4(\sig{\teff{1}})^2 + 0.25(\sig{\fbol})^2 - \sigma_{\fbol\teff{1}}/(\fbol\teff{1})$, which is smaller than our estimate.

Finally, the recovered mass uncertainties are imprecise. Since the dominant source of uncertainty in the masses is the ingress/egress duration (via the primary density), the precision can be improved in two ways. Space-based telescopes such as \emph{Kepler} and {\it TESS} can observe bright transiting systems with significantly better precision than can be achieved from the ground. If observing from the ground instead, then one can decrease the uncertainties by observing multiple transits. 16 (uncorrelated) full transits of similar quality and cadence would yield a factor of 4 improvement in the transit observable uncertainties. In the circular case, the recovered 8\% primary mass uncertainty would be reduced to 5\%, and the companion mass uncertainty would be reduced to 3.8\%. While this is a large number of transits, we note that the large networks of small-aperture telescopes that observe transiting planet candidates can feasibly achieve this quantity; in the case of KELT-12, 15 full and partial transits have been observed \citep{Stevens2017k12}.

\begin{deluxetable}{cccccc}
\tablecaption{Analytic Estimates vs. Actual Precisions Achieved for a Fiducial Transiting Hot Jupiter on Observable and Derived Parameters}
\tablehead{\colhead{Parameter} & \multicolumn{2}{c}{Analytic Value} & \multicolumn{3}{c}{EXOFASTv2 Values}\\
& Circ. & Ecc. & Circ. & Ecc. & Inc.}
\startdata
\sidehead{Transit Parameters}
$\sig{\tau}$ & 8\% & \nodata & 6.7\% & 22\% & 18\% \\
$\sig{\rho}$ & 13\% & 17\% & 6.8\% & 19\% & 18\% \\
$\sig{\delta}$ & 1.1\% & \nodata & 1.7\% & 2.7\% & 3.0\% \\
\sidehead{RV Parameters}
$\sig{K_1}$& 2.2\% & 2.9\% & 1.6\% & 2.8\% & 2.6\%\\
\sidehead{SED Parameters}
$\sig{\teff{1}}$ & \nodata & \nodata & 2.8\% & 2.8\% & 3.0\% \\
$\sig{\pi_p}$ & 0.1\% & 0.1\% & 0.1\% & 0.1\% & 0.1\% \\
\sidehead{System Parameters}
$\sig{M_1}$ & 21\% & 24\% & 8.0\% & 19\% & 19\%\\
$\sig{R_1}$ & 5.6\% & 5.6\% & 1.6\% & 1.7\% & 1.8\% \\
$\sig{M_2}$ & 14\% & 16\% & 4.6\% & 13\% & 13\% \\
$\sig{R_2}$ & 5.6\% & 5.7\% & 1.7\% & 2.1\% & 2.4\%
\enddata
\label{tab:paramcomp}
\end{deluxetable}

\subsection{Other Effects}\label{sec:caveats}
\subsubsection{Second Light}\label{sec:seclight}
We have assumed that the companion contributes negligibly to the total flux of the system, so that we can attribute the total flux solely to the primary star. In reality, the companion does contribute to the total flux, which has two effects on our analysis.

First, neglecting the companion's light leads to an underestimate of the companion-to-primary radius ratio. The transit depth is the difference between the out-of-transit flux $F_{\rm out}$ and in-transit flux $F_{\rm in}$ relative to the out-of-transit flux; for our simplified case, this is
\begin{equation}
    \delta = (F_{\rm out} - F_{\rm in}) / F_{\rm out}.
    \label{eq:blenddepth}
\end{equation}
$F_{\rm out} = F_1 + F_2$ is the total flux from both components and $F_{\rm in} = F_1 + F_2 - k_{\rm true}^2F_1$ is total flux minus the amount of primary stellar flux that is obscured by the companion; $k_{\rm true}$ is the true radius ratio. Writing the flux ratio as $f = F_2/F_1$, and solving for $k_{\rm true}$ in Equation \ref{eq:blenddepth} gives

\begin{equation}
\label{eq:blendratio}
k_{\rm true} = \sqrt{\delta (1+f)},    
\end{equation}

Performing our linear error propagation analysis gives this expression for the fractional uncertainty on $R_2$:
\begin{equation}
\label{eq:}
\Sig{R_2}^2 = \Sig{R_1}^2 + \frac{1}{4}\Sig{\delta}^2 + \frac{1}{4}\Sig{(1+f)}^2,    
\end{equation}

so the uncertainty on $R_1$ remains the dominant contributor to the uncertainty on $R_2$, while the flux ratio and depth uncertainties contribute a smaller amount.

Second, neglecting the companion's light leads to an overestimate of the primary star's bolometric flux in an SED analysis. At fixed distance and \teff, this results in an overestimated primary stellar radius, since $R_1 \propto \sqrt{\fbol}$; this then increases the other primary and companion parameters. Let $f_{\rm bol}$ be the companion-to-primary bolometric flux ratio. The primary's bolometric flux would be overestimated by a factor of $1+f_{\rm bol}$, and Equation \ref{eq:r1pfull} would become

\begin{equation}
    \begin{aligned}
    \Sig{R_1}^2 \approx & 4\Sig{\teff{1}}^2 + \Sig{\pi_p}^2 + \\
    & \frac{1}{4}\Sig{F_{\rm bol,1}}^2 + \frac{1}{4}\left(\frac{\sigma_{f_{\rm bol}}}{1+f_{\rm bol}}\right)^2.
    \label{eq:r1pfull2}
    \end{aligned}
\end{equation}
As before, the precision on \teff{1} dominates the $R_1$ error budget, and the bolometric flux terms are the weakest contributors.

A physically motivated accounting of this increase depends on the wavelength range and stellar parameters in a non-analytic way, so it is beyond the scope of this paper.

\subsubsection{Relative Sizes}
We have also assumed that the companion is small so that $q << 1$ and $k <<1$. We will now consider the effects of a comparably sized companion. We had assumed $q << 1$ to obtain Equation \ref{eq:m2rv} from Equations \ref{eq:massfunc} and \ref{eq:RV}. If we account for a non-negligble mass ratio, then Equation \ref{eq:m2rv} becomes
\begin{equation}
\label{eq:m2rv2} M_2 = (2\pi G)^{-1/3} \frac{K_1P^{1/3}}{\sin i}M_1^{2/3}q^{2/3},
\end{equation}
which shows that the small-$q$ assumption leads to an underestimate of the companion mass. Linear error propagation then yields
\begin{equation}
    \Sig{M_2}^2 \approx \Sig{K_1}^2 + \frac{4}{9}\Sig{M_1}^2 + \frac{4}{9}\left(\frac{\sigma_q}{1+q}\right)^2
\end{equation}

The uncertainty on the RV semi-amplitude dominates the $M_2$ error budget, with the uncertainty on the mass ratio contributing relatively weakly. 

There are two other effects. As $R_2 \rightarrow R_1$, the flux decrease as the companion moves across the limbs of the primary star is no longer linear in time; this is due to the occulted region of the primary star changing nonlinearly during ingress/egress. Thus, the piecewise-linear transit model adopted for our analytic estimates would no longer be valid, as the transit shape would no longer be trapezoidal, even in the absence of limb darkening. In contrast, the Mandel-Agol transit model does account for this effect, albeit while treating both the primary and companion as circular disks.

Moreover, we have used the small-companion assumption to set the companion's contribution to the left-hand side of Equation \ref{eq:densityeq} to zero; however, if the small-companion assumption is false, then the primary stellar density we infer from this equation will be an overestimate.

\subsubsection{Eccentricity}\label{sec:ecc}
\begin{figure*}
\plottwo{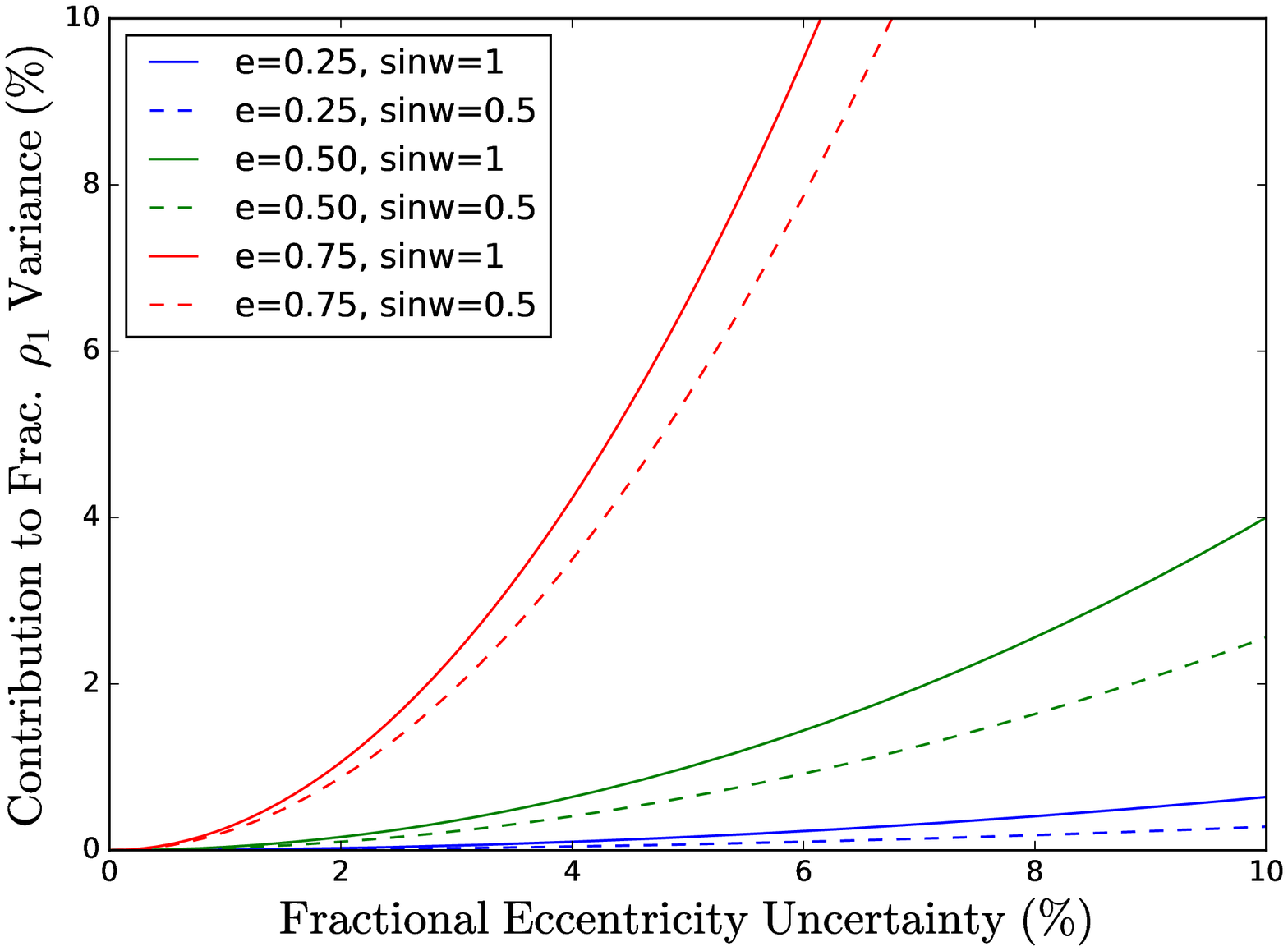}{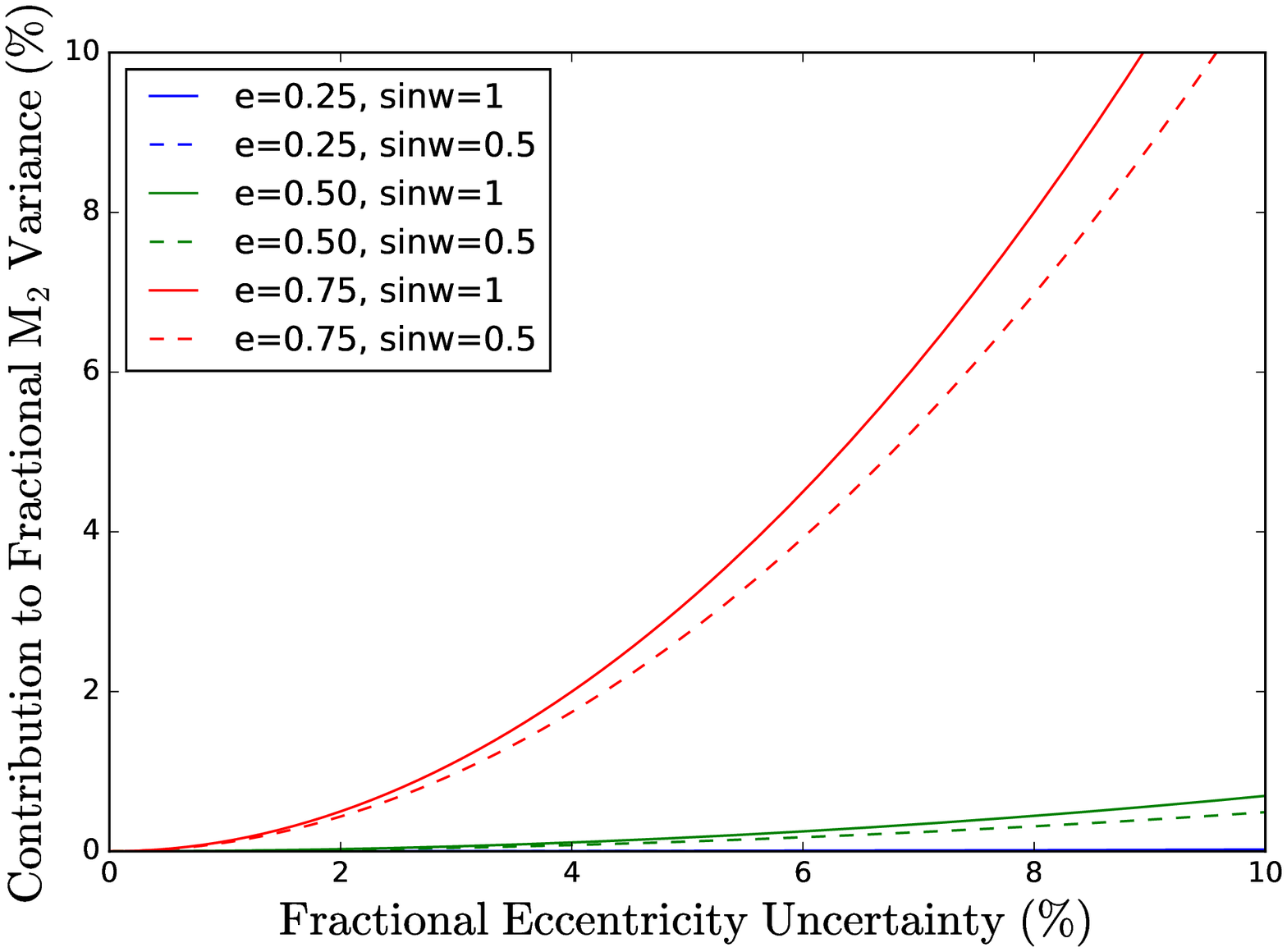}
\caption{\label{fig:ecceffects} Contributions to the squared fractional uncertainty on the primary stellar density (left) and companion mass (right) as a function of the fractional uncertainty on the eccentricity for eccentricity values of 0.25 (blue), 0.5 (green), and 0.75 (red), and $\sin\omega$ values of 1 (solid lines) and 0.5 (dashed lines). Eccentricity uncertainties only modestly contribute to the uncertainty budgets in low-eccentricity cases, while a small eccentricity uncertainty significantly impacts the density and mass uncertainties if the system is highly eccentric.}
\end{figure*}

Nonzero eccentricity complicates the equations for the density and RV uncertainties, among others. For an eccentricity $e$ and argument of periastron $\omega$, the stellar density uncertainty given in Equation \ref{eq:densityeq} becomes (under the assumption that the companion is negligibly small)
\begin{equation}
    \label{eq:densityecc} \rho_1 \approx \rho_{1,circ}\left(\frac{\sqrt{1-e^2}}{1+e\sin\omega}\right)^3,
\end{equation}
and the linear error propagation (ignoring covariances) yields
\begin{equation}
\begin{aligned}
\label{eq:densityeccsigma} \left(\frac{\sigma_{\rho_1}}{\rho_1}\right)^2 \approx & \left(\frac{\sigma_{\rho_1}}{\rho_1}\right)^2_{circ} + 9\left[e(1+e\sin\omega)\sigma_{\sin\omega})\right]^2 \\
& + \left(\frac{3e^2}{1-e^2} + \frac{3e\sin\omega}{1+e\sin\omega}\right)^2\Sig{e}^2. \\
\end{aligned}
\end{equation}
Similarly, the RV semi-amplitude uncertainty becomes
\begin{equation}
\label{eq:Keccsigma} \Sig{K_1}^2 = \Sig{K_1}^2_{circ} + \left(\frac{e^2}{1-e^2}\right)^2\Sig{e}^2.
\end{equation}
In the case of a primary stellar radius from parallax, we can combine Equation \ref{eq:Keccsigma} with Equation \ref{eq:m2pfull} to get the uncertainty on the companion mass in terms of the uncertainty on the eccentricity, the argument of periastron, and the mass in the circular-orbit case:
\begin{equation}
\begin{aligned}
\label{eq:m2eccsigma} \left(\frac{\sigma_{M_2}}{M_2}\right)^2 \approx & \left(\frac{\sigma_{M_2}}{M_2}\right)^2_{circ} + 4\left(\frac{\sigma_{\sin\omega}}{1+e\sin\omega}\right)^2 \\
& + \left(\frac{3e^2}{1-e^2} + \frac{2e\sin\omega}{1+e\sin\omega}\right)^2\Sig{e}^2.
\end{aligned}
\end{equation}
Thus, the fractional eccentricity uncertainty contributes a factor of $3 e^2/(1-e^2)$ to the uncertainties on the primary density and secondary mass, if the argument of periastron is zero. Figure \ref{fig:ecceffects} illustrates the contribution from the eccentricity term to the variances of the primary density and companion mass for different combinations of $e$ and $\sin\omega$ values. While the contributions from the eccentricity uncertainty are small for low-eccentricity systems, they increase dramatically for highly eccentric systems, thus warranting exquisite RV phase coverage and per-point precision.

\subsubsection{Limb Darkening}\label{sec:ld}
To explore the effects of limb darkening on the achievable precision, we consider the same fiducial hot Jupiter as described in Section \ref{sec:HJ} in a circular, edge-on orbit. Along with the simulated RV and SED measurements, we perform \texttt{EXOFASTv2} fits using two different synthetic light curves: one with quadratic limb-darkening coefficients $u_1 = 0.454$ and $u_2 = 0.263$ (corresponding to the $V$ band) and another fit with fixed $u_1 = u_2 = 0$. 

\begin{figure*}
    \centering
    \includegraphics[width=0.33\textwidth]{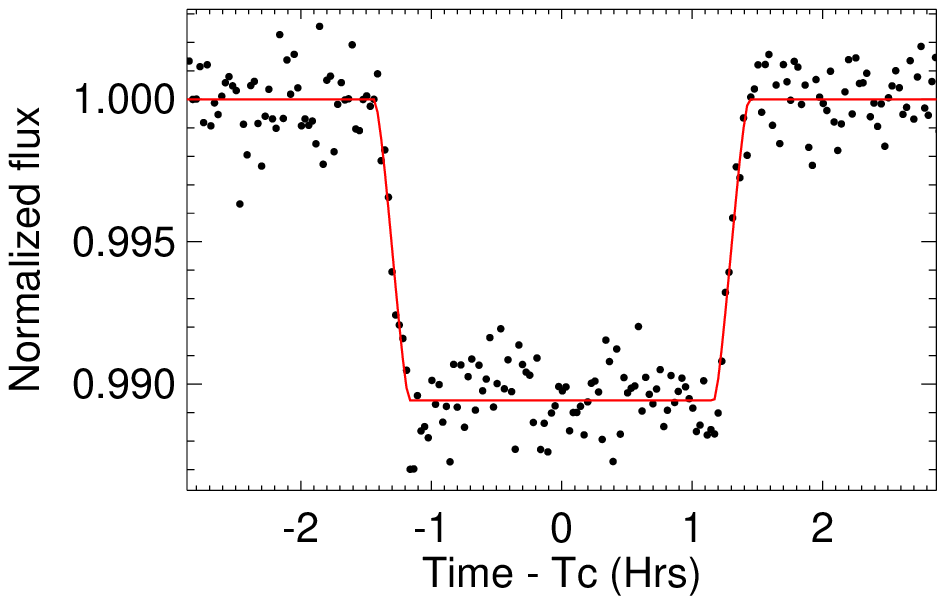}\includegraphics[width=0.33\textwidth]{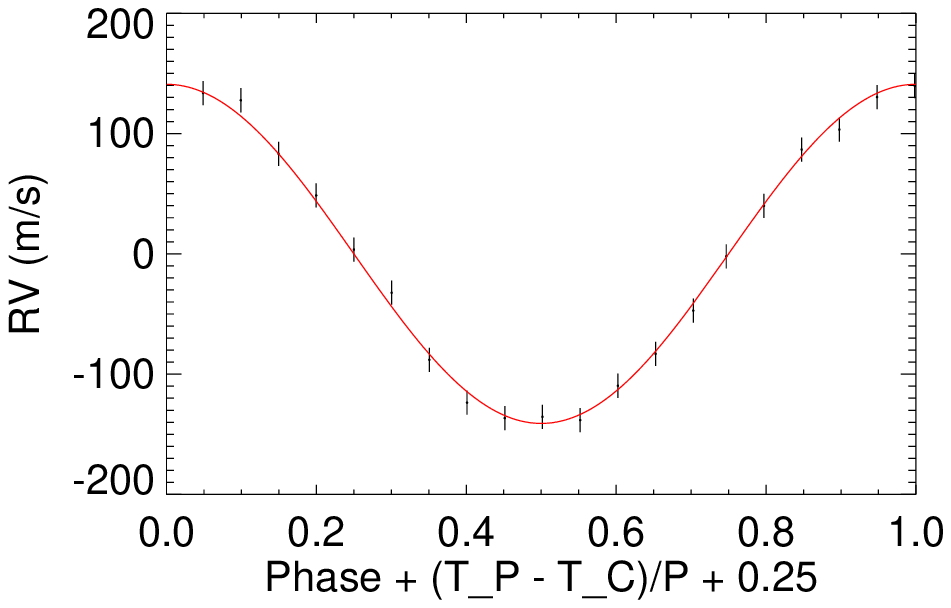}\includegraphics[width=0.33\textwidth]{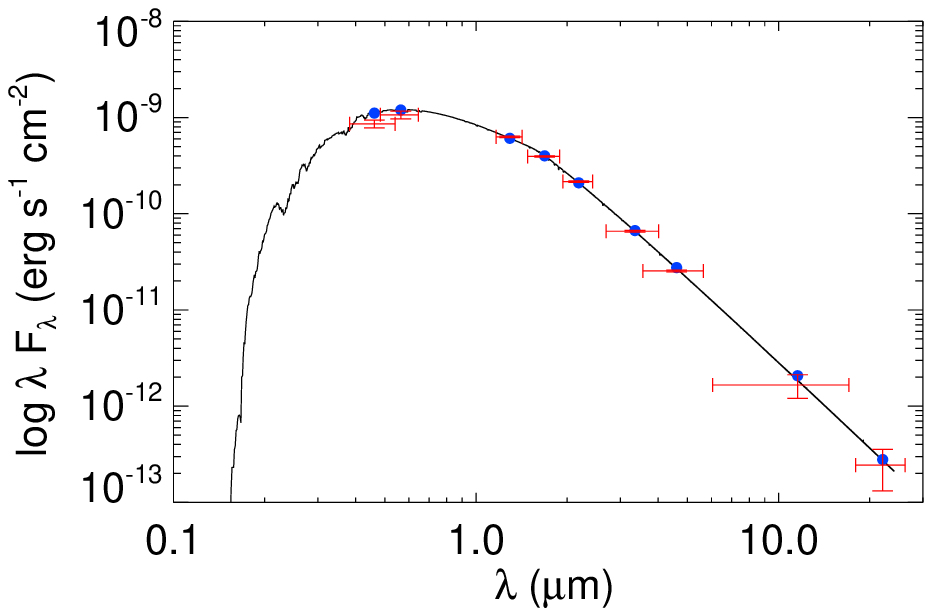}
    \includegraphics[width=0.33\textwidth]{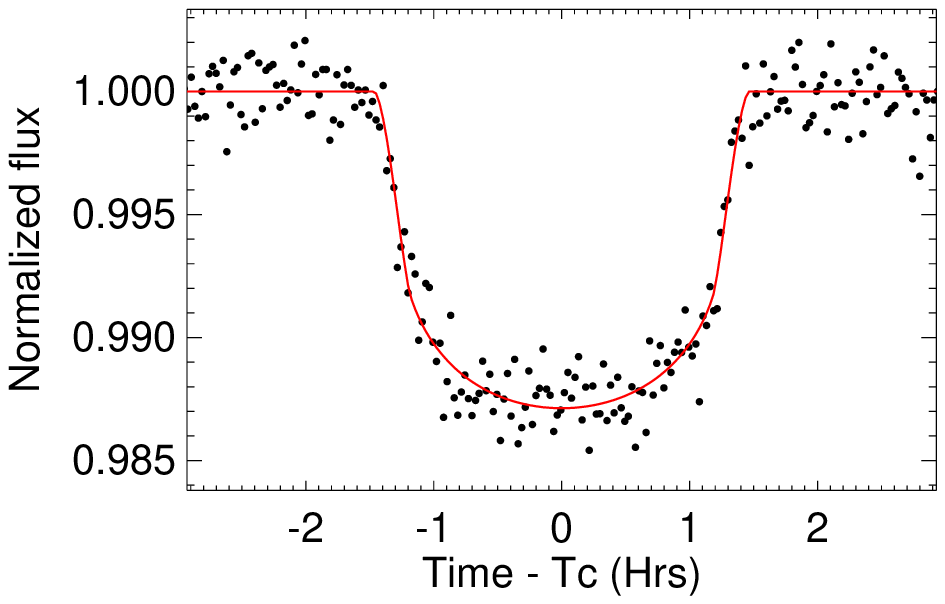}\includegraphics[width=0.33\textwidth]{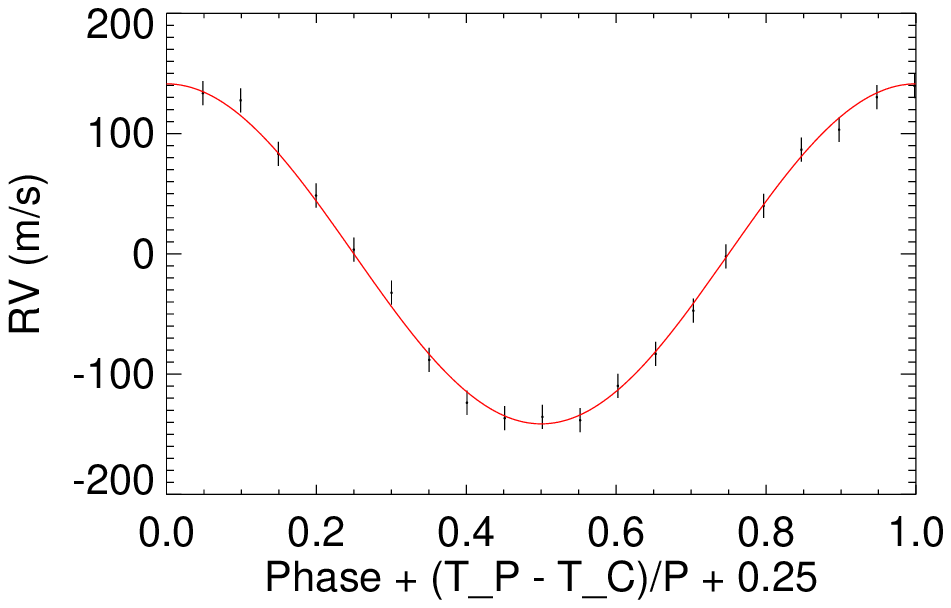}\includegraphics[width=0.33\textwidth]{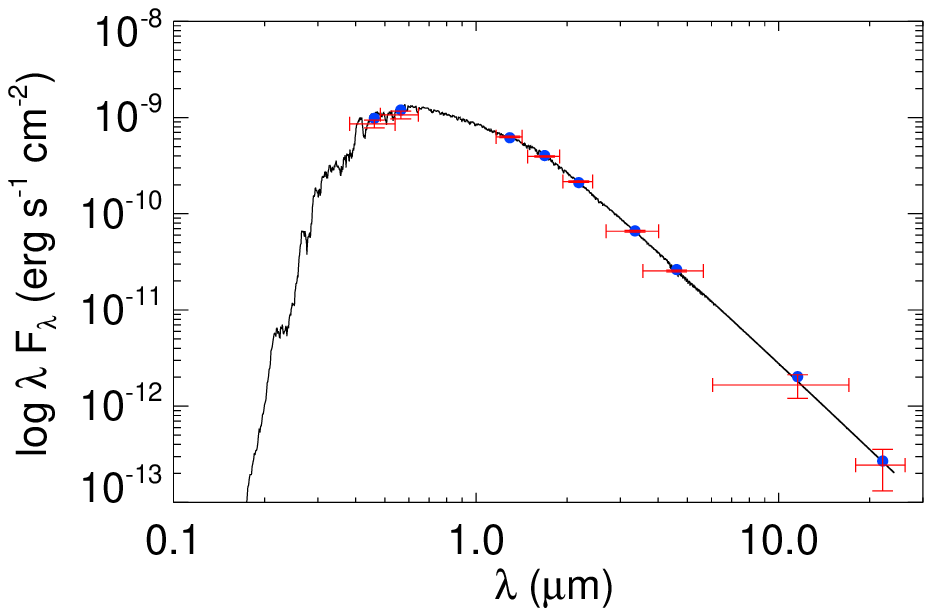}
    \caption{Transits (left), RVs (middle), and SEDs (right) for the same fiducial hot Jupiter transiting a Sun-like star on a circular, edge-on orbit in the limit of no limb darkening (top) and quadratic limb darkening in the $V$ band (bottom). In all cases, the solid lines denote the model fits; in the transit and RV plots, the black dots show the mock data, while the red crosses show the mock broadband fluxes in the SED plot.}
    \label{fig:modelfits_ld}
\end{figure*}

Figure \ref{fig:modelfits_ld} shows the model fits to the light curves, RVs, and SEDs, while Table \ref{tab:paramcompld} gives the uncertainties on the best-fit model parameters for both cases. As expected, the presence of limb darkening increases the uncertainties on the transit ingress/egress duration and depth, which then propagate through to the component masses and companion radius. Interestingly, the effective temperature uncertainty is lower in the presence of limb darkening: at each step in the MCMC chain, the limb-darkening coefficients are selected by interpolating the limb-darkening tables -- in this case, the \citet{Claret2011} tables -- in \teff{1}, \logg, and \feh, so \teff{1} is correlated with the limb-darkening coefficients. These results also suggest that, in the absence of limb darkening, $\sig{\rho_1} \approx 1.5(\sig{\tau})$, while $\sig{\rho_1} \approx \sig{\tau}$ in the presence of limb darkening. The latter relation is the one seen in the fiducial fits of Section \ref{sec:HJ}. This suggests that adding free parameters for limb darkening weakens the strong dependence of the estimate of the density of the primary on the ingress/egress duration, while simultaneously to contributing significantly to the uncertainties on both these quantities.

We note that our analysis involves interpolating in \citet{Claret2011} limb-darkening tables and assuming a quadratic limb-darkening law. As a result, any inaccuracies in the tabulated limb-darkening coefficients -- or any inaccuracy of the underlying limb-darkening equation -- would induce systematic errors. \citet{Espinoza2016} showed how different limb darkening laws result in not only different precision for transit parameters but also in systematic differences between laws at the couple-percent level, which is significant.

Hence, choosing red filters -- such as {\it TESS}' 600--1000 nm bandpass -- will minimize the effects of limb darkening and thus provide the highest precision on transit/eclipse observables while also improving the accuracy of the inferred transit parameters.

\begin{deluxetable}{ccc}
\tablecaption{Uncertainties on Fiducial Transiting Hot Jupiter Parameters with and without Limb Darkening}
\tablehead{\colhead{Parameter} & \colhead{No Limb Darkening} & \colhead{Limb Darkening}}
\startdata
\sidehead{Transit Parameters}
$\sig{\tau}$ & 4.6\% & 23\% \\
$\sig{\rho}$ & 6.3\% & 23\% \\
$\sig{\delta}$ & 1.3\% & 5.2\% \\
\sidehead{RV Parameters}
$\sig{K_1}$& 1.6\% & 1.5\%\\
\sidehead{SED Parameters}
$\sig{\teff{1}}$ & 4.7\% & 2.6\% \\
\sidehead{System Parameters}
$\sig{M_1}$ & 12\% & 22\% \\
$\sig{R_1}$ & 3.2\% & 1.7\% \\
$\sig{M_2}$ & 8.2\% & 16\% \\
$\sig{R_2}$ & 3.3\% & 3.1\% \\
\enddata
\label{tab:paramcompld}
\end{deluxetable}

\section{Discussion}\label{sec:yield}

There already exists a large number of known single-lined eclipsing systems for which we can measure precise and accurate stellar and planetary parameters. Single-lined EBs are a common false positive in exoplanet transit surveys \citep{Collins2018}, and the KELT and HATNet surveys in particular have identified nearly 200 such systems covering a range of orbital periods -- including periods longer than 10 days, for which only a couple M-dwarf binaries are known -- and spectral types. Only a couple of eclipsing systems containing M dwarfs are known with such long orbital periods, such as the 41-day period M-M binary LSPM J1112+7626 \citep{Irwin2011}; by measuring masses and radii for the M dwarfs in the single-lined EBs, we will be able to obtain precise masses and radii in a regime where binary physics (e.g. tidal interactions) should be negligible.
In addition, the {\it TESS} survey is expected to find nearly 300,000 EBs \citep{Sullivan2015,Sullivan2017}. A large number of those should be single-lined EBs for which {\it TESS} light curves should provide stellar densities to a precision of order $1\%$ or better, e.g. 0.25\% in the case of KELT-11 \citep{Beatty2017}.

Additionally, it should be possible to measure surface gravities from granulation-tracing brightness variations (flicker; \citealt{Bastien2016}) to 0.1 dex from {\it TESS}' 30-minute cadence light curves; from the 2-minute cadence light curves, both flicker and the timescale of granulation and acoustic oscillations (\citealt{Kallinger2016}) should be measurable, with the latter producing gravities as precise as 4\%. These techniques should be applicable to any star showing surface convection, i.e., stars below the Kraft \citep{Kraft1967} break, including late F-, G-, and K-star primaries of single-lined EBs. Such measurements would provide a second set of stellar parameters measured not only for the same systems (the single-lined EBs) but from the same light curves, providing both a consistent comparison sample of stellar parameters and improved precision on the stellar parameters, assuming the parameters derived from flicker/granulation agree reasonably with those from the light curve, RV, and SED modeling \citep[see][]{StassunGrav:2017}.

Given {\it TESS}' sensitivity to bright ($V \lesssim 12$) stars, there will be an exceptionally large number of EBs with precise photometry and systematics-limited \emph{Gaia} parallaxes \citep{deBruijne2012} that will require only precise photometric and RV follow-up. Moreover, \emph{Gaia} spectrophotometry from 330 to 1050 nm for each star will provide excellent measurements around the SED peaks for F, G, and K stars. If the proposed all-sky {\it SPHEREx} \citep{Dore2016,Dore2018} satellite launches in the early 2020s, then it will provide spectrophotometric coverage from approximately 500 to 5000 nm; such coverage will capture the SED peaks for stars of spectral types mid-K and later and the Rayleigh-Jeans tail for stars of earlier spectral type. 

Additionally, the combination of \emph{Gaia} and {\it SPHEREx} spectrophotometry with broadband photometry in the literature will allow for direct extinction measurements, as well as measurements of nearly all the flux for bright stars, which will almost completely break the dependence on stellar model atmospheres (and thus on \teff, which is a fitted SED parameter) for measuring the bolometric flux and thus make these measurements much closer to ``empirical.''

As mentioned in Section \ref{sec:setup}, the companion in a single-lined EB usually contributes more to the total flux from the system in NIR wavelengths. In favorable cases, it will be possible to measure the RV orbit of the companion in the NIR by obtaining spectra around the orbital phases of maximum and minimum RV amplitude (which can be known from measuring the primary star's RV orbit in optical filters) and performing a two-dimensional cross-correlation analysis. In this way, \citet{Bender2012} measured the NIR RV orbit for the M-dwarf companion in the inner K-M binary of the Kepler-16 circumbinary planet system \citep{Doyle2011} and measured the binary components' masses to $<3\%$. This approach can serve as a crucial test of systematics in the joint analysis of the eclipse photometry, primary star's RVs, and the SED.

This leads to one potentially promising way of addressing the accuracy of \teff\ scales, if such a single-lined EB has measured transits/eclipses; RV orbits of both components; full photometric SED coverage from {\it Gaia}, {\it SPHEREx}, and others; and a {\it Gaia} parallax. With radii measured from analysis of the eclipses and both RV orbits and a luminosity calculated from the parallax distance and the directly summed and dereddened bolometric flux, a \teff can be calculated and compared to \teff\ values from other sources.

Furthermore, single-lined EBs provide an opportunity to measure accurate metallicities and abundances of the primary star, since the total flux (and thus the spectrum) of the system is dominated by the primary star. The companion's metallicity and abundance can then be assumed to be similar to those of the primary (although in some cases, it may be possible to verify this assumption by direct observations). In the DEBCat catalog of well-studied detached EBs \citep{Southworth2015}, of the 15 systems containing M dwarfs, only three  -- the M-M EBs PTFEB 132.707+19.810 (\citealt{Gillen2017, Kraus2017}) and CM Dra (\citealt{Morales2009,Terrien2012}), plus the G-M binary V530 Ori \citep{Torres2014} -- have measured metallicities. Single-lined EBs thus have the potential to increase the number of M dwarfs in EBs with metallicities significantly. 

Finally, we note that there will be some single-lined EBs for which it will be possible to obtain multiple complementary constraints of the primary star's physical properties, e.g. from the SED, spectroscopy, and asteroseismology and/or granulation-driven variability. These systems will serve as particularly important benchmarks with which to compare the stellar masses and radii recovered from these independent methods and test and refine the empirical relations used to constrain the stellar properties (i.e., the asteroseismological or flicker scaling relationships), and they are likely to be the best-characterized, ``Rosetta stone'' systems as a result.

\section{Summary and Conclusions}\label{sec:summary}
By applying simple linear error propagation techniques and ignoring covariances, we have derived analytic expressions for the uncertainties on the masses and radii of a single-lined eclipsing EBs under the simplifying assumptions that the companion is dark, small, and on a circular orbit around a primary star that exhibits no limb darkening. We have shown how uncertainties in different measured quantities -- the transit ingress/egress duration, a spectroscopic surface gravity, a parallax and SED parameters, and asteroseismic quantities -- affect the uncertainties on the masses and radii. We showed that orbital eccentricity contributes appreciably to the uncertainties on the primary stellar density and companion mass for highly eccentric systems. We also demonstrated how the presence of limb darkening increases the uncertainties on the transit observables and can decrease the uncertainty on the primary star's effective temperature (and thus its contribution to the primary star's radius uncertainty) if one adopts a prior on the limb-darkening coefficients based on the properties of the primary star (e.g. $\teff{1}$ and $\logg$) in the light-curve fit.

We determined the constraints that isochrones and empirical \citet{Torres2010} relations place on the mass and radius of a star: $M \propto ZR^{\alpha}$, where $\alpha \approx 0.2$ for the Torres relations and $\alpha \approx 0.36$ for the isochrones. These constraints are fairly complementary to the constraint from a density, $M \propto \rho R^3$, or surface gravity, $M \propto 10^{\log g}R^2$. For both cases, we showed how the uncertainty on $Z$ propagates through to the mass and radius uncertainties when combined with a stellar density measurement; in the case of the Torres relations, we also showed how the $Z$ and its uncertainty depend on effective temperature and metallicity values and uncertainties.

By generating a simulated light curve, RV orbit, and broadband fluxes for a fiducial hot Jupiter transiting a Sun-like star, we have argued that it will be possible to measure precise, accurate, and nearly model-independent (i.e. without using stellar evolution models or empirically calibrated relations) masses and radii for single-lined EBs in the {\it Gaia} era; there still is, however, a dependence on transit and/or eclipse models, as well as a nominal dependence on stellar atmospheres for SED modeling. As discussed in detail by \citet{StassunGaiaEB:2016} and by \citet{StassunGaiaPlanets:2016}, even the bolometric flux obtained from the SED fitting is itself essentially model-independent, thanks to the availability of broadband fluxes that now span the entire SED from $\sim$0.1 to $\sim$20 $\mu m$ for most stars of interest in the sky. With the thousands to hundred-thousands of EBs expected from {\it TESS}, it will be possible to increase the sample of EBs with precisely known masses, radii, effective temperatures, surface gravities, metallicities, abundances, and Galactic positions and velocities; as such, these systems can become benchmark systems for a multitude of planetary, stellar, and Galactic investigations.

\acknowledgements\label{sec:ack}
We thank J. Eastman and J. Rodriguez for assistance with an early version of \texttt{EXOFASTv2}. We also thank M. Pinsonneault, J. Johnson, G. Somers, and J. Tayar for discussions pertaining to asteroseismology and R. Narayan and S. Villanueva for insightful discussions related to stellar SEDs/spectroscopy and transit parameters, respectively.

Work by BSG and DJS was partially supported by NSF CAREER grant AST-1056524.
\bibliography{biblio.bib}
\end{document}